\documentclass[11pt]{article}
\pdfoutput=1
\usepackage{jheppub}
\usepackage{amsthm}

\usepackage{hyperref}

\usepackage{xcolor}
\usepackage{mathtools}
\usepackage{tensor}
\usepackage{empheq}

\usepackage{tikz}
\usepackage{graphicx}
\usepackage{subfigure}
\usetikzlibrary{calc,fadings,decorations.pathreplacing}
\usepackage{verbatim}
\usepackage[normalem]{ulem}
\usepackage{braket}
\usepackage{jheppub}

\preprint{LITP-25-14}

\title{Quantum-Corrected Evaporation and Absorption Cross-Section  of Near-Extremal Rotating Black Holes }

\author[a,b]{Shu Luo}

\author[b]{ and Leopoldo A. Pando Zayas}

\emailAdd{ls040629@mail.ustc.edu.cn, lpandoz@umich.edu}

\affiliation[a]{University of Science and Technology of China, Hefei, Anhui 230026, China }

\affiliation[b]{Leinweber Institute for Theoretical Physics, 
University of Michigan, Ann Arbor, MI 48109, USA}

\abstract{ We revisit the Hawking evaporation history of low-temperature rotating black holes by taking into consideration the strong quantum fluctuations known to be present in the near-horizon, near-$\mathrm{AdS_2}$ throat region governed by an effective action that includes Schwarzian and two gauge modes. Imposing compatibility of this quantum framework with the semiclassical results creates a novel link of the black hole angular momentum and electric charge before and after emission, leading to a nontrivial interplay among the superradiance effect, eigenstate thermalization hypothesis and microscopic statistic description. We evaluate single scalar (neutral and charged) emission of Kerr-Newman and single and di-particle emission of photons, gravitons and spinors in the Kerr spacetime. We uncover that quantum corrections may affect late time evaporation rates, which further slows down the whole evaporation process because of the near-balance between the $s$-wave channel and the superradiance channel. Specifically, we find energy decay of the form $E(t)\sim t^{-8/21}$ for neutral scalar emission of a small, slowly rotating and charged black hole which differs from the analogous spherically symmetric quantum correction $E(t)\sim t^{-2/5}$ already suppressed with respect to the semiclassical rate $E(t)\sim t^{-1}$. We also discuss the quantum cross section for rotating black holes and point out various new features.}

\begin{document}

\maketitle

\section{Introduction}
Black holes are ideal laboratories in which to test  concepts on  the path to constructing a theory of quantum gravity. Near-extremal black holes are particularly suitable as probe of quantum gravity because of the expectation that at low temperatures quantum effects can dominate the dynamics. The central role of quantum effects in near-extremal black holes was first inferred more than thirty years ago when it was identified that thermodynamic notions break down for near-extremal black holes at extremely low temperatures~\cite{Preskill:1991tb}. More elaborate forms of this statement were presented  over the years~\cite{Maldacena:1998uz, Page:2000dk}. Essentially, at very low-temperatures the emission of one single Hawking quantum completely destroys the thermodynamic approximation.

There has recently been an improved understanding of the dynamics taking place in the throat of near-extremal black holes. The progress was originally achieved in the simple framework of the two-dimensional Jackiw-Teitelboim (JT) gravity~\cite{Jackiw:1984je, Teitelboim:1983ux}. This model provides a controlled quantum setup  that allows to describe strong gravitational quantum effects~\cite{Maldacena:2016upp, Jensen:2016pah}. This understanding of the nature of the quantum fluctuations was later applied to higher-dimensional near-extremal black holes and led to important modifications to the low-temperature thermodynamics of black holes~\cite{Iliesiu:2020qvm, Heydeman:2020hhw, Boruch:2022tno, Iliesiu:2022onk}. Further work has confirmed that the same mechanism applies to rotating black holes~\cite{Kapec:2023ruw,Rakic:2023vhv,Maulik:2024dwq} and even to certain asymptotically de-Sitter black holes~\cite{Blacker:2025zca, Maulik:2025phe,Arnaudo:2025btbb}.

The role of the quantum fluctuations in the throat has been investigated beyond thermodynamics and into the realm of more dynamical questions such as their influence on Hawking radiation~\cite{Brown:2024ajk,Maulik:2025hax,Lin:2025woff,Bhattacharjee:2025wfv,Rakic:2025svg} and broader aspects of the quantum cross section~\cite{Emparan:2025sao,Biggs:2025nzs,Emparan:2025qqff,Betzios:2025sctt,Jiang:2025cyl}; there have also been some studies exploring implications for aspects of the AdS/CFT correspondence~\cite{Daguerre:2023cyx,Liu:2024gxr,Liu:2024qnh,Nian:2025oei,PandoZayas:2025snm,Cremonini:2025yqe,Gouteraux:2025exs,Kanargias:2025vul}. One key argument made by the authors of~\cite{Brown:2024ajk} is that, as opposed to the semiclassical treatment, the Hawking quanta frequency is bounded above by the microcanonical energy of the initial state; essentially, stating that the black hole can not emit more energy than its own initial mass. The quantum effects considered in~\cite{Brown:2024ajk} lead to the whole evaporation rate being slower than the semiclassical prediction, while some semiclassical statements still hold in the presence of quantum corrections. For example, Page's argument that the main channel for single particle emission is the one with angular quantum number $l$ as small as possible is still valid. In the quantum regime, photons and gravitons can only be emitted via di-particle channels~\cite{Brown:2024ajk,Lin:2025woff}.

The universal semiclassical result stating that the cross section of the $s$-wave scattering of a black hole is proportional to its area has long provided a bridge between a hypothetical scattering experiment and the microstates of the black hole~\cite{Das:1996we}. Recently, this connection has been revisited in the context of the quantum corrections typical of near-extremal charged black holes~\cite{Emparan:2025sao, Biggs:2025nzs, Emparan:2025qqff,Betzios:2025sct}.  Another analysis was presented in~\cite{Biggs:2025nzs} who followed the evaporation as a process that does not necessarily fit in a fixed thermodynamic ensemble. Recently, some connections between the quantum absorption cross section and the shear viscosity of holographic fluids at very low temperatures in the context of the AdS/CFT correspondence have also been put forward~\cite{Nian:2025oei,PandoZayas:2025snm,Cremonini:2025yqe,Gouteraux:2025exs,Kanargias:2025vul}. 

In this manuscript, we consider Hawking evaporation of rotating charged black holes and obtain results that are partly different from the spherically symmetric case reported in previous literature. We extend the analysis of~\cite{Brown:2024ajk} to a situation where the black hole can exchange angular momentum as well as electric charge via  emission of particles, thus leading to a much richer space of possibilities. Most importantly, we find that compatibility of the quantum result with semiclassical ones imposes charge conservation, given that one takes the correct statistical description together with eigenstate thermalization hypothesis (ETH). There are some remarkable new phenomenons. First, the dominant channel can only be single particle emission even in the quantum regime, a situation that is not {\it a priori} clear. Second, for the case of neutral scalar particle emitting from a relatively small and rotating black hole, the tendency of superradiance effect overwhelming the emission will lead to a near balance cancellation between the $s$-wave channel (which decreases BH local energy or equivalently energy over extremity) and the superradiance channel (which {\it increases} BH local energy). Therefore, the late time evaporation rate will become even slower than expected in the quantum corrected spherically symmetric case.

Apart from evaporation history, we also generalize some of the recent results about quantum cross section to the case of rotating black holes. We confirm that there is a window of quantum transparency for photon and graviton scattering. We also discuss the enlargement of superradiance effects when one approaches the quantum regime. For scalars, we verify that the low frequency enhancement of the absorption as well as high frequency universality also make sense for rotating black holes.

The rest of the manuscript is organized as follows. We briefly review the low-temperature limit of near-extremal rotating black holes in Section \ref{Sec:Review}. In Section \ref{Sec:Framework} we present the framework for introducing quantum fluctuations in the throat in the case of rotating black holes. Section \ref{Sec:EmissionRates} discusses emission rates of scalar, photon, graviton and spinor as well as a discussion of di-particle emission rates. In Section \ref{Sec:Evaporation} we study the evaporation history of rotating black holes. In Section \ref{Sec:CrossSection} the quantum cross section and its difference from the  semiclassical predictions is discussed. We present some concluding remarks and point to interesting open questions in Section \ref{Sec:Discussion}.

\section{ Classical thermodynamics of near-extremal rotating black holes}\label{Sec:Review}

We start by considering the following Kerr-Newman background:
\begin{equation}
    \mathrm{d}s^2=-\frac{\Delta}{\rho^2}(\mathrm{d}t-a\sin^2\theta\mathrm{d}\phi)^2+\frac{\rho^2}{\Delta}\mathrm{d}r^2+\rho^2\mathrm{d}\theta^2
    +\frac{\sin^2\theta}{\rho^2}[a\mathrm{d}t-(r^2+a^2)\mathrm{d}\phi]^2\,
\end{equation}
where 
\begin{eqnarray}
    \rho^2=r^2+a^2\cos^2\theta\,,\qquad \Delta=r^2-2mr+a^2+e^2=(r-r_{+})(r-r_{-})\,,
\end{eqnarray}
and
\begin{equation}
    A=-\frac{er}{r^2+a^2\cos^2\theta}(\mathrm{d}t+a\sin^2\theta\mathrm{d}\phi)\,.
\end{equation}

From the parameters contained in the metric, we have thermodynamic variables in the following form (we have used the unit in which $[G_N]=-2$ and $[M]=1$, so $[r_{+}]=-1$ and e.t.c.):
\begin{equation}
    T_{H}=\frac{r_+-r_-}{4\pi(r_+^2+a^2)}\,,\qquad S=\frac{\pi}{G_N}(r_+^2+a^2)\,,
\end{equation}
\begin{equation}\label{01}
    \Phi=\frac{Qr_+}{r_+^2+a^2}\,,\qquad \Omega_H=\frac{a}{r_+^2+a^2}\,,
\end{equation}
which satisfies the first law
\begin{equation}
    \delta M=T_H\delta S +\Omega_H\delta J+\Phi_H\delta Q\,.
\end{equation}
Now we aim to perform low temperature expansion of Kerr-Newman black hole. In the canonical ensemble, we need to expand all parameters in $T_{H}$ fixing charges, $J$ and $Q$. Let us suppose
\begin{equation}\label{02}
r_{+}=r_0+a_1T_{H}+a_2T_{H}^2+O(T_{H}^3)\,,\quad r_{-}=r_0+b_1 T_{H}+b_2 T_{H}^2+O(T_{H}^3)\,,
\end{equation}
and bring the expression into 
\begin{equation}
    T_{H}=\frac{r_+-r_-}{4\pi(r_+^2+\frac{4J^2G_N^2}{(r_{+}+r_-)^2})}\,,
\end{equation}
\begin{equation}
    r_+r_-=\frac{4J^2G_N^2}{(r_++r_-)^2}+G_NQ^2\,,
\end{equation}
compare the coefficients of $T_{H}^n$ on both sides is enough to confirm
\begin{equation}
    a_1=2\pi(r_0^2+a_0^2)=-b_1\,,
\end{equation}
\begin{equation}
    a_2=10\pi^2 r_0(r_0^2+a_0^2)\,,\qquad b_2=-6\pi^2 r_0(r_0^2+a_0^2)\,,
\end{equation}
Thus 
\begin{equation}\label{06}
    M=\frac{r_0}{G_N}+2\pi^2 \frac{r_0(r_0^2+a_0^2)}{G_N}T_{H}^2\,, \qquad S=\frac{\pi}{G_N}(r_0^2+a_0^2)+4\pi^2\frac{r_0(r_0^2+a_0^2)}{G_N}T_{H}\,,
\end{equation}
and the latter gives the characteristic breaking energy scale at which semiclassical methods no longer make sense
\begin{equation}\label{Eq:C}
    E_{brk}^{-1}=C=\frac{1}{4\pi^2}\Big{(}\frac{\partial S}{\partial T_H}\Big{)}\Big{|}_{T_H \rightarrow 0}=\frac{r_0(r_0^2+a_0^2)}{G_N}\,.
\end{equation}
This expansion is generally well-known one, with energy only quadratically relies on temperature, which directly predicts the breakdown of semiclassical treatment in extremal case.

There are considerations of mainly focusing on canonical ensemble when discussing quantum correction of near-extremal rotating black holes \cite{Rakic:2023vhv}. The key point is that the contractable ``time" circle in rotating black holes with Euclidean signature (and those with correct periodicity condition for fermions compatible with spin structure) is a mixture of $t$ and $\varphi$, and so the partition function is computing $\mathrm{Tr}(e^{-\beta H+\beta\Omega J})$ for microstates. But then the chemical potential related to rotation will have the periodicity $i\beta\Omega\rightarrow i\beta \Omega+2\pi$ for pure bosons and $i\beta\Omega\rightarrow i\beta\Omega+4\pi$ for involving fermions, or in other words we can only fix the chemical potential modulo $2\pi i T_H/4\pi iT_H$. However Kerr metric itself does not respect such symmetry, so in principle one has to sum over all different gravitational saddle points. Though in BPS case these generally does not contribute to the index due to Fermionic zero modes~\cite{Iliesiu:2021are}, for non-supersymmetric case there is no similar argument to rule them out.

Still if we restrict the discussion to Lorentz signature, one can fix both chemical potentials $\Omega$ and $\Phi$ on the real axis. Then with the expressions in Eq.(\ref{01}) $a(r_{+})=\frac{1}{2}(\frac{1}{\Omega}\pm \sqrt{\frac{1}{\Omega^2}-4r_+^2})$ and $Q(r_+)=\frac{\Phi_H [r_{+}^2+a(r_+)^2]}{r_+}$ and still bring the ansatz \ref{02} into two key relations $T_H=\frac{r_+-r_-}{4\pi(r_+^2+a(r_+)^2)}$ and $r_+r_-=a(r_+)^2+G_NQ(r_+)^2$, it is enough to confirm all coefficients:
\begin{equation}
    r_{+}=r_0-2\pi (r_0^2+a_0^2)\frac{r_0^2(r_0^2-a_0^2)}{a_0^2(3r_0^2-a_0^2)}T_H+O(T_H^2)\,,
\end{equation}
\begin{equation}
    r_{-}=r_0-2\pi (r_0^2+a_0^2)\frac{r_0^4+5a_0^2r_0^2-2a_0^4}{a_0^2(3r_0^2-a_0^2)}T_H+O(T_H^2)\,,
\end{equation}
so the charges are given by 
\begin{equation}
    M=\frac{r_0}{G_N}-\frac{2\pi}{G_N}  (r_0^2+a_0^2) \frac{r_0^4+2a_0^2r_0^2-a_0^4}{a_0^2(3r_0^2-a_0^2)} T_H+O(T_H^2)\,,
\end{equation}
\begin{equation}
J=Ma=M_0a_0-\frac{2\pi}{G_N}\frac{(r_0^2+a_0^2)^2}{a_0}T_{H}+O(T_H^2)\,.
\end{equation}
\begin{equation}
Q=Q_0-2\pi Q_0\frac{r_0(r_0^2+a_0^2)^2}{a_0^2(3r_0^2-a_0^2)}T_H+O(T_H^2)\,.
\end{equation}
\begin{equation}
    S=\frac{\pi}{G_N}(r_0^2+a_0^2)-\frac{4\pi^2}{G_N}\frac{r_0^3(r_0^2+a_0^2)^2}{a_0^2(3r_0^2-a_0^2)}T_H+O(T_H^2)\,,
\end{equation}
and as a verification we can confirm that the expansion shown above can satisfy Smarr relation (equivalently first law)
\begin{eqnarray}
    M=2T_HS+2\Omega_HJ+\Phi_HQ\,,
\end{eqnarray}
at the order $T_{H}$. We need to emphasize that the accurate physical meaning of the expansion above, however, is still unclear on statistical mechanical level, and it is unlikely to relate them to a dual quantum theory partition function computation. The main concerns comes from, for instance, the negative heat capacity leading to an unstable system, and the nonvanishing linear $T_H$ coefficient of mass parameter, which does not accord to usual circumstance in semiclasssical discussion. 

\section{Quantum Corrected Hawking Evaporation}\label{Sec:Framework}

This section reviews the framework used to include quantum fluctuations that are strong in the throat region of near-extremal Kerr-Newman black holes. The conceptual framework was first utilized in \cite{Brown:2024ajk} elaborating on ideas described in \cite{Mertens:2019bvy}. We follow the implementation of this framework as applied to rotating black holes first described in \cite{Maulik:2025hax} but also improve on some significant aspects left unresolved there.

\subsection{Effective quantum theory of the throat}\label{Subsec:EffectiveQTheory}
Near-extremal black holes in four dimensions have a long $\mathrm{AdS_2}$ throat in the near horizon region. Any perturbation that enters this region, be it scalar, vector or tensor will naturally interact with the boundary gravitational modes and other gauge modes describing the very low energy theory living in the throat.  We model these interactions following a version of the AdS/CFT correspondence. Namely, any mode that penetrates the near-horizon region is described by an operator whose two-point function is determined by the low-energy theory. When computing emission rates, this two-point function enters via Fermi's golden rule. Alternatively, one can consider the gravitational action as effectively reducing to a JT/Schwarzian theory plus a couple of gauge modes at very low energies. The low energy effective action is completely determined by a few couplings and it takes the general form \cite{Davison:2016ngz,Mertens:2019tcm,Iliesiu:2020qvm,Iliesiu:2022onk}: 
\begin{eqnarray}\label{sch}
    I_{eff}&=&-C\int_{0}^{\frac{1}{T_{H}}}d\tau \{\tan(\pi T_{H}f(\tau)),\tau\}-\frac{K_e}{2}\int_{0}^{\frac{1}{T_H}}d\tau (\partial\tau\phi_e+i(2\pi \mathcal{E}_eT_H)\partial_\tau f)^2\nonumber \\
    &-&\frac{K_r}{2}\int_{0}^{\frac{1}{T_H}}d\tau (\partial\tau\phi_r+i(2\pi \mathcal{E}_rT_H)\partial_\tau f)^2\,.
\end{eqnarray}
There are two $U(1)$ gauge fields that accompany the Schwarzian action. One $U(1)$ comes from the higher-dimensional electric field and the other $U(1)$ is from  what would have been $SO(3)$ in the RN case. This latter $U(1)$ originates in the fact that the near-horizon extremal Kerr solution has an enhanced isometry $SL(2,R)\times U(1)$ where $U(1)$ is the unbroken subgroup, $U(1) \sim SO(2)\subset SO(3)$. 

 For the RN black hole all of these modes are pseudo-Goldstone modes that belong to massless KK modes corresponding to two-dimensional reduced action~\cite{Iliesiu:2020qvm}. The reduction itself is not possible in the presence of  rotation. In this case there are some other dynamical fields, apart from dilaton, that couple to the pseudo-Goldstone modes~~\cite{Castro:2019crn,Castro:2021csm,Castro:2021fhc,Castro:2025itb}. However, for the quantum aspects we are mostly concerned in this manuscript we make the simplifying assumption that, to leading order, the effects of those couplings might be neglected. Our argument relies in certain universality that has been established regarding the nature of tensor zero modes in the higher-dimensional theories we seek to approximate. In higher dimensions, the zero modes that eventually generate the Schwarzian action are in the kernel of the  Lichnerowicz operator; this was first discussed for Kerr black holes in asymptotically flat four-dimensional  spacetime in~\cite{Kapec:2023ruw,Rakic:2023vhv} and further extended to asymptotically AdS black holes in various dimensions in \cite{Maulik:2024dwq}. More recently, this universality has been rigorously proved in \cite{PandoZayas:2026vbg} for black holes in asymptotically flat, AdS and dS and dimensions four, five and six.

We understand that the higher-dimensional zero modes may still couple to other modes such as those discussed in~~\cite{Castro:2019crn,Castro:2021csm,Castro:2021fhc,Castro:2025itb}, however, in the effective action paradigm, the existence of zero modes provides strong justification for the action in  Eq.(\ref{sch}) as a starting point. In particular, such action is compatible with the higher dimensional derivations of one-loop corrected black hole thermodynamics ~\cite{Iliesiu:2020qvm,Rakic:2023vhv,Maulik:2024dwq,PandoZayas:2025snm}:
\begin{eqnarray}
    S_{1-\mathrm{loop}}\sim \frac{3}{2}\log \frac{T}{T_0}\,,
\end{eqnarray}
Let us also highlight that in the original application to RN black holes ~\cite{Iliesiu:2020qvm}, where dimensional-reduction is available, the Schwarzian action is to be rigorously interpreted as an effective field theory, given that  the framework requires a finite cutoff JT theory. It is natural to assume this framework in the rotating case. Formally speaking, the theory space we focus on is governed by large diffeomorphisms of $\mathrm{nAdS_2}$, rotation around $z$ axis and gauge field that is constant in the  transverse space:
\begin{eqnarray}
    \mathcal{M}=\mathcal{M}_{Sch}\times \mathcal{M}_{U(1)}^2\,,
\end{eqnarray}
\begin{eqnarray}
    \mathcal{M}_{Sch}=\mathrm{Diff}(\mathbf{S^1})/\mathrm{PSL}(2,\mathbb{R})\,, \qquad \mathcal{M}_{U(1)}=\mathrm{Loop}(U(1))/U(1)\,.
\end{eqnarray}

All five parameters in the above low energy effective action are determined  by the higher-dimensional  thermodynamic variables
\begin{equation}
    C=\frac{1}{4\pi^2}\Big{(}\frac{\partial S}{\partial T}\Big{)}_{Q,J}\Big{|}_{T_{H}\rightarrow 0}=\frac{r_0(r_0^2+a_0^2)}{G_N}\,,
\end{equation}
\begin{equation}
    K_e=-\Big{(}\frac{\partial Q}{\partial \Phi_H}\Big{)}_{T_{H},\Omega_H}\Big{|}_{T_{H}\rightarrow 0}=\frac{r_0(r_0^2+a_0^2)(3a_0^2-r_0^2)}{a_0^2(a_0^2-3r_0^2)}\,,
\end{equation}
\begin{equation}
    K_r=-\Big{(}\frac{\partial J}{\partial \Omega_H}\Big{)}_{T_{H},\Phi_H}\Big{|}_{T_{H}\rightarrow 0}=\frac{2r_0(r_0^2+a_0^2)}{G_N}\,. \label{Eq:Kr}
\end{equation}
\begin{equation}
    \mathcal{E}_e=\frac{1}{2\pi}\Big{(}\frac{\partial S}{\partial Q}\Big{)}_{T_H,J}\Big{|}_{T_H\rightarrow 0}=\frac{1}{2\pi}\Big{(}\frac{\partial S_0}{\partial Q}\Big{)}_{J}=\frac{G_NQ_0^3}{a_0^2+r_0^2}\,,
\end{equation}
\begin{equation}
    \mathcal{E}_r=\frac{1}{2\pi}\Big{(}\frac{\partial S}{\partial J}\Big{)}_{T_H,Q}\Big{|}_{T_H\rightarrow 0}=\frac{1}{2\pi}\Big{(}\frac{\partial S_0}{\partial J}\Big{)}_{Q}=\frac{2a_0r_0}{a_0^2+r_0^2}\,,
\end{equation}
where $C=E_{brk}^{-1}$ is the scale of conformal breaking, and there is a relation $\mathcal{E}=-\frac{\beta}{2\pi}(\mu-\mu_0),$ as $T_H\rightarrow 0$. Here, as only fluctuations above extremity can be captured by the Schwarzian theory, we can simply take $\mu_0=0$. We denote by $K$ the charge susceptibility or compressibility.

Just as we mentioned in Sec.\ref{Sec:Review}, the grand canonical ensemble in rotating spacetime might not have a very clean definition at the  statistical level. It is beneficial to mention that the case we discuss in the last section (to totally fix the chemical potential with regard to temperature changing) is merely a special case of what we discuss here, in which $\mathcal{E}$ as a linear order coefficient of $T_H$ is fixed at a particular value, so it is clear that not all of $\mathcal{E}$ can guarantee a well-defined and convergent partition function. This accords to the general conclusion about asymptotically flat black holes in~\cite{Iliesiu:2020qvm}. We still refer to what has been done in RN-(AdS) spacetime to get a speculative expression in analogy, whose correctness will be proved step by step through our deduction. The one-loop partition function is given by the Schwarzian effective theory as a function of the inverse temperature, $\beta$, and chemical potential, $\mu$, with contributions from different charges:
\begin{equation}
    Z(\beta,\{\mu_i\})=Z_{\mathrm{sch}}(\beta,C)Z_{U(1)}(\frac{\beta}{K},\beta \mu)=e^{S_0}\Big{(}\frac{2\pi C}{\beta}\Big{)}^{\frac{3}{2}}e^{\frac{2\pi^2C}{\beta}}\sum_{n\in \mathbf{Z}}e^{-\beta(\frac{n^2}{2K}-\mu n)}\,.
\end{equation}
The first factor above, $e^{S_0}$, is the semiclassical ground state entropy, the second and third factors are the contribution of the one-loop Schwarzian and  the leading order semiclassical contribution, respectively. Finally, the sum comes from the $U(1)$ mode, $Z_{U(1)}$. One can interpret the $U(1)$ partition function more broadly as a character over the Hilbert space $Z_{U(1)}=\mathrm{tr}(e^{-\beta (H-\mu Q)})$ where $H=\frac{Q^2}{2K}$ is the energy and $Q$ is the  charge. The above expression is a special case of the JT partition function coupled to general gauge fields,  
\begin{eqnarray}
    Z_{G}=\frac{1}{\mathrm{Vol}(G)}\sum_{R}\mathrm{dim}(R)\chi_R(e^{\beta\mu})e^{-\beta\frac{C_2(R)}{2K}}\,,
\end{eqnarray}
in which $R$ is a certain representation and $\chi_R(V)$ is the character in $R$ for a group element $V\in G$ while $C_2(R)$ is its quadric Casimir.

We now let $q=-n$, and then by an inverse Laplace transformation \cite{Mertens:2022irh}
\begin{equation}\label{density par}
    Z(\beta,\mu)=\int dE \,Z(\beta,\mu,E)=\int dE \sum_{q}\rho(E,q) e^{-\beta E-\beta\mu q}\,,
\end{equation}
one can get the density of states with energy $E$ and charge $q$
\begin{equation}\label{03}
    \rho(E,q)=\frac{C}{2\pi^2}e^{S_0}\sinh\left(2\pi\sqrt{2C(E-\frac{q^2}{2K})}\right)\,,
\end{equation}
here $E$ is the energy above extremity and $q$ has the meaning of gauge charge above extremity. If there are contributions from both the electric charge and angular momentum charge, as in the case of the Kerr-Newman black hole, one can naturally extend the above expression to include both charges in an additive way. We note that in near-extremal RN or BPS black holes similar expressions of density of states make sense for any fixed gauge charge channel, regardless of whether the grand canonical ensemble itself can be well-defined. Moreover, we will mainly focus on microcanonical or canonical ensemble in the following computation, so we can confidently adopt Eq.(\ref{03}) as a reasonable starting point.  

Still there are various issues in quantitatively connecting the actual fluctuating physical charge to this $U(1)$ charge that appears in the density of states. While the general relation is unclear, we note that $q$ is an integer and its value should be altered through the emission of a particle with nonzero $z$-spin. Specifically, it should be $q_i-q_f=m$ for a particle emission in channel $m$. A hint that this is possible comes from the comparison between the semiclassical result (when $\zeta\equiv E_i/E_{brk}=CE\rightarrow \infty$) and the quantum one computed in a reasonably selected microcanonical ensemble that can {\it effectively} describe the whole physical emission process . 

Let us illustrate this idea in more detail. We now focus on the energy flux involving a fixed energy but fluctuated charge, i.e., a ``partially" grandcanonical ensemble with given $E_i$ and chemical potential $\beta,\mu$:
\begin{eqnarray}\label{Eq:GC}
   \frac{d E}{dt}\Big{|}_{\mathrm{total}}= Z(\beta,\mu,E_i)^{-1}\sum_{q_i}\rho(E_i,q_i)e^{-\beta\mu q_i-\beta E_i}\frac{dE}{dt}\Big{|}_{\mathrm{micro},E_i,q_i}\,.
\end{eqnarray}
Assuming $dE/dt$ depends rather smoothly on $q_i$ (which will be shown later in our computation), and considering $\beta$ is very large, we only study the saddle point of the weight factor above $f(q_i)=\rho(E_i,q_i)e^{-\beta\mu q_i}$, which is located at $q_{0}$ satisfying
\begin{eqnarray}\label{saddle}
    \tanh(2\pi\sqrt{2C(E-\frac{q_{0}^2}{2K})})=\frac{Cq_{0}}{\mathcal{E}K\sqrt{2C(E-\frac{q_{0}^2}{2K})}}\,.
\end{eqnarray}
In the semiclassical limit the left side can be approximated as unity, leading to 
\begin{eqnarray}\label{peak}
    q_{0}=\frac{\sqrt{8\zeta}\mathcal{E}}{\sqrt{1+2\mathcal{E}^2}}\gg 1\,,
\end{eqnarray}
where we have used the explicit expressions for $K$ and $C$ given in Equations \eqref{Eq:Kr} and \eqref{Eq:C}, respectively and implying $K/C=2$.  We can further estimate the variance to be very small: 
\begin{eqnarray}\label{Eq:width}
    \frac{\Delta q}{q_{0}}=\frac{1}{q_{0}}\sqrt{\frac{-2f(q_{0})}{f''(q_{0})}}\approx \sqrt{\frac{1}{2\pi\mathcal{E}^2\sqrt{2(1+2\mathcal{E}^2)\sqrt{\zeta}}}}\rightarrow 0\,.
\end{eqnarray}
Therefore, we can approximate the whole contribution of the energy flux with given $E_i$ to all come from that of $q_{0}$ in the semiclassical limit. This determines the correct effective microcanonical channel for our later discussion when comparing with the semiclassical regime. Crucially, we will show in Sec.~\ref{scalar} that this approximation, together with the requirement of $U(1)$ charge conservation $q_i-q_f=m$, can exactly give the correct answer for the match between the quantum-corrected emission and the semiclassical results at least for the  ``near-superradiance" frequency region (or as we will later show, the {\it bona fide} Hawking radiation region). We will use that the semiclassical physics is described by ETH temperature $T_H=\sqrt{\frac{E_{brk}}{2\pi^2}(E_i-\frac{q_i^2}{2K})}$~\cite{Lam_2018,Brown:2024ajk}.

\subsection{Coupling to matter in the bulk}\label{Subsec:Coupling}
The crucial new ingredient, compared to the standard computation of black hole emission, is that we couple the external perturbation to the low-energy quantum theory characterized by the Schwarzian and two $U(1)$ modes. The free field expansion of the bulk scalar field at asymptotic infinity is 
\begin{equation}\label{infty}
\hat\phi(t,r,\theta,\varphi)=\sum_{lm}\int_0^{\infty} d\omega \frac{e^{-i\omega t}}{\sqrt{4\pi\omega}r}[\hat{a}_{\omega lm}e^{-i\omega r}+\hat{b}_{\omega lm}e^{i\omega r}]Y_{lm}(\theta,\varphi)+\mathrm{h.c.}\,,
\end{equation}
where $\hat{a}_{\omega lm}$ and $\hat{b}_{\omega lm}$ are the ingoing and outgoing modes annihilation operators respectively, and from the classical solution one has $\hat{b}_{\omega lm}=-e^{-il\pi}\hat{a}_{\omega lm}$. Inverting the above expression \eqref{infty}, we have 
\begin{equation}
    \hat{a}_{\omega lm}=\sqrt{\frac{\omega}{4\pi}}\int rdr d\Omega Y_{lm}^{*}[\hat{\phi}(0,r,\theta,\varphi)+\frac{i}{\omega}\dot{\hat{\phi}}(0,r,\theta,\varphi)]\,,
\end{equation}
so that 
\begin{equation}
    [\hat{a}_{\omega' l'm'},\hat{a}_{\omega lm}^{\dagger}]=\delta_{l'l}\delta_{m'm}\delta(\omega'-\omega)\,.
\end{equation}
The classical solution of $\hat\phi_{\omega lm}$ in the bulk can be written as
\begin{equation}
    \hat{\phi}_{\omega lm}(t,u)=\hat{\phi}_0 u^{1-\Delta}+O(u^{\Delta})\,,\; u\rightarrow 0\,,
\end{equation}
where $u$ is the Poincare coordinate in the AdS$_2$ region. In AdS/CFT correspondence the coefficient of nonnormanizable part of the bulk field as it approaches the conformal boundary is regarded as the source term interacting with the boundary operator $\mathcal{O}_{lm}$, and normalizable one is the response. In near extremal limit, the near horizon region and far region get decoupled and the quantum tunneling out of horizon vanishes, so the normalizable coefficient. Then,  we have
\begin{equation}
\hat{\phi}_0=\mathcal{N}_{lm}\omega^{\Delta-\frac{1}{2}}\hat{a}_{\omega lm}\propto \hat{c}_{\omega lm}\,,
\end{equation}
where $\hat{c}_{\omega lm}$ is obtained from $\hat{a}_{\omega lm}$ in the overlap region through matching with the far region, which has the same functional relationship of $\omega$ as $\hat{\phi}_0$. Here, $\mathcal{N}_{lm}$ is a universal normalization factor in a fixed channel still to be determined through matching the quantum results with the semiclassical ones in the large $E_i$ regime. Thus, the coupling Hamiltonian can be written as 
\begin{equation}
H_I=\sum_{lm}\mathcal{O}_{lm}\mathcal{N}_{lm}\int_0^{\infty}d\omega \omega^{\Delta-\frac{1}{2}}(\hat{a}_{\omega lm}+\hat{a}_{\omega lm}^{\dagger})\,.
\end{equation}

In the above construction we need to make the transformation $\omega\rightarrow\omega_{\rm eff}= \omega-m\Omega_H-e\Phi$ for any process relevant with particle absorption or emission in NHR, including the calculation of the bosonic factor in the semiclassical and matrix element as well as the density function in quantum computation. Note that $\omega_{\rm eff}$ here appears because the local energy of emission near the horizon is different from that observed at conformal infinity. In other words we are working in the Frolov-Thorne vacuum~\cite{Frolov:1989jh}, equivalently observing it in a zero angular momentum frame (ZAMO), which plays the role of uniformly accelerating Rindler observer near horizon. This is a natural requirement for recovering the semiclassical results which also respect the principle of equivalence. The above transformation of the frequency is equivalent to 
\begin{equation}
    \phi\rightarrow \phi'=\phi e^{ie\Phi},\qquad \varphi \rightarrow \varphi'=\varphi-\Omega_H t\,.
\end{equation}
Now focusing on rotation, for the $m>0$ channel (we will explain later while the $m<0$ channel is generally not relevant in our discussion) we reexpress Eq.(\ref{infty}) as 
\begin{eqnarray}
\begin{aligned}
\hat\phi(t,r,\theta,\varphi)=\sum_{lm}\int_0^{\infty} d\omega_{\rm eff} \frac{e^{-i\omega_{\rm eff} t}}{\sqrt{4\pi\omega}r}[\hat{a}_{\omega_{\rm \rm eff} lm}e^{-i\omega r}+\hat{b}_{\omega_{\rm eff} lm}e^{i\omega r}]Y_{lm}(\theta,\varphi')\\
    +\int_{-m\Omega}^{0} d\omega_{\rm eff} \frac{e^{i(-\omega_{\rm eff}) t}}{\sqrt{4\pi\omega}r}[\hat{a}^{\dagger}_{-\omega_{\rm eff}, lm}e^{-i\omega r}+\hat{b}^{\dagger}_{-\omega_{\rm eff}, lm}e^{i\omega r}]Y_{lm}(\theta,\varphi')+\mathrm{h.c.}\,,
\end{aligned}
\end{eqnarray}
and the second line explicitly shows that in the superradiance region the whole physical image of the process near the black hole is altered: one should not regard the ``radiation" as a Hawking quanta with local energy $\omega_{\rm eff}<0$ tunneling through the barrier and escaping to infinity with energy $\omega$. It is more appropriate, instead, to consider one quantum with local energy $-\omega_{\rm eff}$ being absorbed by black hole inside the barrier while the observer at infinity receives a Hawking quantum with energy $\omega$. Either way, we can effectively treat it from the infinity point of view.

In the quantum treatment of RN black hole, the dual operator $\mathcal{O}_{lm}$ is a spherical tensor of type $(l,m)$ and we can use the Wigner-Eckhart theorem to obtain its matrix element between two arbitrary $SO(3)$ states. On the CFT side, it is an operator with conformal dimension $\Delta$, whose value can be read off from the power behavior of the corresponding bulk scalar field. Here as we only encode the axial degree of freedom into the effective theory, in the case when there is no electric $U(1)$ charge transfer, we can only shift the ground level and treat the matrix element between energy sectors with $q_i=q_f=0$. This can be obtained through direct computation of the two point correlation function $<\mathcal{O}(\tau_1)\mathcal{O}({\tau_2})>$ in the gravity side through path integral (only take disk topology into consideration)~\cite{Yang:2018gdb} and 
\begin{equation}\label{correlator}
    <\mathcal{O}(\tau_1)\mathcal{O}({\tau_2})>=\mathrm{Tr}[e^{-\beta H}\mathcal{O}(\tau_1)\mathcal{O}(\tau_2)]=\int dE_1dE_2 e^{-\beta E_2}e^{-\tau (E_1-E_2)}
\rho(E_1)\rho(E_2)|\bra{E_1}\mathcal{O}\ket{E_2}|^2\,.
\end{equation}
so that we can read out the matrix element directly. Another way to derive this is through defining two sided quasi-micro canonical state $\ket{E}_G=\frac{1}{\delta E}\sum_{|E_n-E|\leq \delta E}\ket{E_n}_L\ket{E_n}_R\approx \frac{N_n}{\delta E}\ket{E_n}_L\ket{E_n}_R$ and by introducing intermediate geodesic length states
\begin{equation}\label{07}
\begin{aligned}
|\bra{E_f}\mathcal{O}\ket{E_i}|^2\approx\frac{\delta E^2}{N_iN_f}{}_G\bra {E_f}O_LO_R\ket{E_i}_G=
    \int d\lambda\frac{{}_G\braket{E_f|\lambda} e^{-\lambda\Delta}\braket{\lambda|E_i}_G}{\rho_{E_i}\rho_{E_f}}\\
    =2e^{-S_0}\frac{\Gamma(\Delta\pm i\sqrt{2CE_f}\pm i\sqrt{2CE_i})}{(2C)^{2\Delta}\Gamma(2\Delta)}\,.
\end{aligned}
\end{equation}
Including a $U(1)$ charge exchange labeled by initial and final states, we analogously  have  
\begin{eqnarray}
|\bra{E_f,q_f}\mathcal{O}_{lm}\ket{E_i,q_i}|^2=2e^{-S_0}\frac{\Gamma(\Delta\pm i\sqrt{2C(E_f-q_f^2/{2K})}\pm i\sqrt{2C(E_i-q_i^2/2K)})}{(2C)^{2\Delta}\Gamma(2\Delta)}\delta_{q_f,q_i-m}\,.
\end{eqnarray}
as both $q_i$ and $m$ can be regarded as marking the representation of the $SO(2)\sim U(1)$ gauge field originated from the spin around $z$ axis.

Now that we have obtained the expression for both the density of states and the matrix element, we can use time-dependent perturbation theory in quantum mechanics by Fermi's golden rule in the Frolov-Thorne vacuum which leads to
\begin{equation}\label{05}
    \Gamma_{i\rightarrow f}=2\pi |\bra{\omega}\hat{\phi}_0\ket{0}|^2|\bra{E_f,q_f}\mathcal{O}_{lm}\ket{E_i,q_i}|^2\delta(E_f+\omega_{eff}-E_i)\,.
\end{equation}
The total transition rate is 
\begin{equation}\label{transition rate}
    \Gamma_{\mathrm{total}}=\int d\omega \Gamma_{\mathrm{em}}(\omega)=\int d\omega\int dE_f \rho(E_f,q_f)\Gamma_{i\rightarrow f}\,,
\end{equation}
and the energy flux in the microcanonical ensemble in the state $\ket{E_i,q_i}$ is given by
\begin{equation}
    \frac{d E}{dt d\omega}=\Gamma_{\mathrm{em}}(\omega)\omega=\omega \int_{0}^{\infty}d E_{f}\rho(Ef,q_f)\Gamma_{i\rightarrow f}\,,
\end{equation}
and the delta function imposes before and after evaporation there are
\begin{equation}\label{Eq:exchange}
    E_f+\omega_{\rm eff}=E_i\,, \qquad q_f+m=q_i\,.
\end{equation}
We will bring all expressions into this formula and get the explicit result in Section~\ref{scalar}.

\section{Emission Rate of Fields with Various Spins}\label{Sec:EmissionRates}

Having discussed the general framework for introducing the effects of quantum corrections in the throat region of near-extremal black holes during Hawking evaporation in Section~\ref{Sec:Framework}, we now turn to the details of the emission rates for  various particles. While there are some  results in~\cite{Maulik:2025hax,Rakic:2025svg}, here we address some issues left previously unresolved or ignored. In particular, we present a consistent treatment of the photon and graviton emission by addressing issues with their associated effective conformal dimensions.

\subsection{Emission rate of massless scalar}\label{scalar}
One prerequisite for fully specifying the quantum flux formula is determining the greybody factor. To achieve this goal let us start by considering the radial part of the scalar equation (with electric charge $e$ and angular quantum numbers $l,m$) in the Kerr-Newman spacetime
\begin{equation}\label{teukolsky}
    \frac{d}{dr}(\Delta\frac{dR_0}{dr})+\Big{(}\frac{((r^2+a^2)\omega-am-eQr)^2}{\Delta}+2am\omega-K_{lm}\Big{)}R_0=0\,.
\end{equation}
The equation for the angular part is ($y=\cos\theta$)
\begin{eqnarray}
    \frac{d}{dy}((1-y^2)\frac{d S_s}{dy})-\Big{(}a^2\omega^2(1-y^2)+\frac{m^2}{1-y^2}-K_{lm}\Big{)}S_s=0\,.
\end{eqnarray}
By introducing $x=\frac{r-r_+}{r_+-r_-}=(r-r_{+})/\delta$ as a dimensionless coordinate ranging from [0,$\infty$] (when $\delta=0$ this treatment breaks down, and the formula will be generally different from the non-extremal case), we can transform the radial equation  into the form 
\begin{equation}\label{Eq:radial_a}
    x(x+1)R_0''+(2x+1)R_0'+\Big{(}\frac{p_{-1}x^4+p_0x^3+p_1x^2+p_2x+p_3}{x(x+1)}-p_4\Big{)}R_0=0\,,
\end{equation}
 with 
\begin{equation}
    p_{-1}=\delta^2\omega^2\,, \qquad p_0=2\omega\delta(2r_+\omega-eQ)\,,
\end{equation}
\begin{equation}
    p_1=e^2Q^2-6eQr_{+}\omega+2\omega(-am+a^2\omega+3r_+^2\omega)\,,
\end{equation}
\begin{equation}
    p_2=\frac{2}{\delta}(eQ-2r_+\omega)(am+eQr_+-a^2\omega-r_+^2\omega)\,,
\end{equation}
\begin{equation}
    p_3=\frac{(am+eQr_+-a^2\omega-r_+^2\omega)^2}{\delta^2}\,,\qquad p_4=K_{lm}-2am\omega\,.
\end{equation}
In near horizon limit we can neglect the $x^4$ and $x^3$ terms, keeping only the latter three terms in the last term of Eq. \eqref{Eq:radial_a}. To get an analytical expression we adopt the ansatz $R_0=x^{h_1}(x+1)^{h_2}\chi$, which leads to 
\begin{equation}
\begin{aligned}
    x(x+1)\chi''+[(1+2h_2)x+(1+2h_1)(x+1)]\chi'+\qquad \qquad \qquad\\
    \{\frac{(h_1^2+p_3)+[(h_1+h_2)(1+2h_1)+p_2]+[(h_1+h_2)^2+h_1+h_2+p_1]x^2}{x(x+1)}-p
    _4\}\chi=0\,,
\end{aligned}
\end{equation}
and by choosing $h_1=-i\sqrt{p_3}$ and $h_2=i\sqrt{p_1-p_2+p_3}$, we get a standard form of the hyper-geometric equation: 
\begin{equation}
    x(x+1)\chi''+[(2h_1+2h_2+2)x+1+2h_1]\chi'+[(h_1+h_2)(1+2h_1)+p_2-p_4]\chi=0\,.
\end{equation}
Thus, we have the purely-inner solution as 
\begin{equation}\label{04}
    R_0=x^{-i\sqrt{p_3}}(x+1)^{i\sqrt{p_1-p_2+p_3}}F(\frac{1}{2}-i\sqrt{p_3}+i\sqrt{p_1-p_2+p_3}-\tilde{\beta},\frac{1}{2}-i\sqrt{p_3}+i\sqrt{p_1-p_2+p_3}+\tilde{\beta},1-2i\sqrt{p_3},-x)\,,
\end{equation}
where $\tilde{\beta}=\sqrt{\frac{1}{4}-p_1+p_4}$. Note that in the above solution we have chosen a convenient boundary condition because the greybody factor is the same for the inner and correct mixed boundary condition of quantum fluctuation in which the normalizable coefficient vanishes.  

\begin{figure}[htbp]
	\centering
\includegraphics[width=0.8\textwidth]{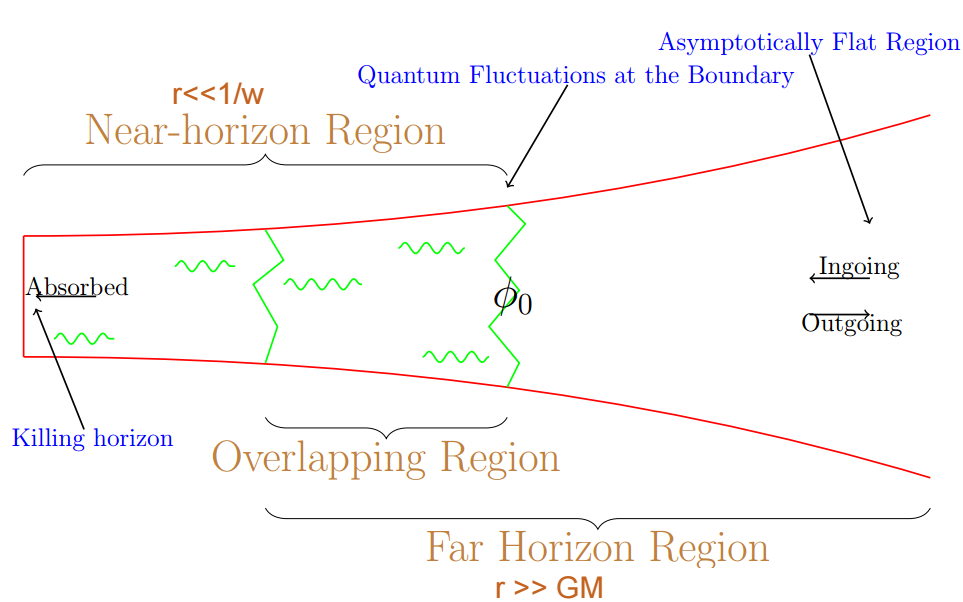}
\caption{Schematic depiction of the black hole absorption process leading to Hawking radiation as
spontaneous emission. We emphasize the role of quantum fluctuations in the throat region in green.}
\label{fig:00}
\end{figure}

This solution is valid until one reaches the outmost $\mathrm{AdS_2}$ throat, where we expect there is an overlapping region for this solution and the asymptotic behavior at infinity. Such expansions are needed to determine the greybody factor, which requires conditions $p_{-1},p_0 \ll p_1,p_2,p_3$, or equivalently $(r_+-r_-)\omega\ll1$. This is a necessary condition for all later analytical computations that we will conduct. The overlapping region is just the intersection of the ``far" and ``near" region $r_0\ll r\ll 1/\omega$, see Fig.\ref{fig:00} for an illustration. 
In this region we can approximate Eq.(\ref{04}) as  
\begin{equation}\label{general}
\begin{aligned}
    R_0\rightarrow x^{-\frac{1}{2}+\tilde{\beta}}\frac{\Gamma(1-2i\sqrt{p_3})\Gamma(2\tilde{\beta})}{\Gamma(\frac{1}{2}-i\sqrt{p_3}-i\sqrt{p_1-p_2+p_3}+\tilde{\beta})\Gamma(\frac{1}{2}-i\sqrt{p_3}+i\sqrt{p_1-p_2+p_3}+\tilde{\beta})}\\+x^{-\frac{1}{2}-\tilde{\beta}}\frac{\Gamma(1-2i\sqrt{p_3})\Gamma(-2\tilde{\beta})}{\Gamma(\frac{1}{2}-i\sqrt{p_3}-i\sqrt{p_1-p_2+p_3}-\tilde{\beta})\Gamma(\frac{1}{2}-i\sqrt{p_3}+i\sqrt{p_1-p_2+p_3}-\tilde{\beta})}\,, \;1\ll x\ll \delta^{-1}\,,
\end{aligned}
\end{equation}
and by introducing an effective angular quantum number $\tilde{l}$ as $\tilde{l}=\tilde{\beta}-\frac{1}{2}$, we rewrite it as
\begin{equation}
    R_0 \rightarrow b_1 x^{\tilde{l}}+b_2x^{-\tilde{l}-1}\,.
\end{equation}
Following the AdS/CFT dictionary, we take the unnormalized source term in dual CFT as $\phi_0\propto b_1$, and the conformal dimension of the dual operator to be $\Delta=\tilde{l}+1$. 

Before implementing the matching method, let us clarify the conditions for computing the greybody factor and the central role of the effective conformal dimension for neutral particles. We now claim that for physically relevant situations we can make the following approximations. First of all we are working in low frequency region, {\it i.e.}, (i) $\delta\omega\ll 1$. Then we face two situations: (ii) for the $m=0$ channel, we have $\omega r_0\ll 1, \omega a_0\ll1,$ etc. (iii) for the  $m\neq 0$ channel and in near-superradiance region, we have $\omega_{\rm eff}r_0\ll1,$ etc, where $\omega_{\rm eff}=\omega-m\Omega_H$.

Let us clarify the validity of the approximation above for universal channels. One would fear these conditions break down for sufficiently large values of the frequency. In that regime, however, Hawking's semiclassical computation is already a good approximation. Note also that an unbounded frequency, $\omega$, would lead to an imaginary conformal dimension in the effective theory pointing also to the breakdown of our approximation. However, there is no such problem if we set a high energy cutoff because the matrix element in Eq.~(\ref{07}) vanishes for $\omega_{eff}=E_i$, so the frequency is restricted to the range $-m\Omega_H<\omega_{eff}<E_i$. The existence of such a high energy cutoff is not surprising at all since the near horizon $\mathrm{AdS}_2$ geometry as well as the holographic duality framework are themselves valid only in the near-extremal limit. Increasing the energy above extremality by considerable amounts prevents meeting such conditions required to  utilize our effective theory. We thus assume $E_i=\zeta E_{brk}$, in which $\zeta$ may be large but has to be not too large (at least $\zeta\ll G_NM^2$). We will call this the ``semiclassical limit", which is actually still an overlapping region. 

In the $m=0$ channel, recall that the complete expression for $E_{brk}$ is
\begin{equation}
    E_{brk}=\frac{G_N}{r_0(r_0^2+a_0^2)}=\frac{M}{(G_NM^2)^2(1+k^2)}\ll M, \qquad k=\frac{a_0}{r_0}\,,
\end{equation}
as $G_NM^2=(M/m_{pl})^2\approx 10^{76}$ for stellar mass black holes. Therefore, if we take the effective temperature as given by the eigenvalue thermalization hypothesis~\cite{Lam_2018}, we have
\begin{equation}\label{1011}
   \delta\omega=\frac{\omega}{E_i}\frac{\sqrt{8\zeta^3}}{(G_NM^2)^2(1+k^2)}\ll \omega r_0=\frac{\omega}{E_i}\frac{\zeta}{(G_NM^2)(1+k^2)}\ll 1\,.
\end{equation}
and similarly $\omega a_0\ll 1$. The combination $\beta \omega$, however, needs not necessarily be small
\begin{equation}
    \beta \omega=\sqrt{\frac{2\pi^2}{E_iE_{brk}}}\omega=\frac{\omega}{E_i}\sqrt{2\pi^2\zeta}\in[0,\sqrt{2\pi^2\zeta}]\,,
\end{equation}
and can, in fact, be very large especially in the ``semiclassical limit". 

In the $m\neq 0$ channels, we need to make the transformation $\omega \rightarrow \omega_{eff}=\omega-m\Omega_H$ in the calculation of Schwarzian matrix element on the quantum side and bosonic factor on the semiclassical side, but we are still particularly interested in the region where $\omega_{eff}\in[0,E_i]$. We call this the near-superradiance region in frequency domain and this region is most physically important as it is regarded as the ``real" Hawking radiation with positive local energy Hawking quanta, where we might have to abandon the approximation that $\omega r_0\ll 1$. Actually, replacing $\omega$ in Eq.(\ref{1011}) by $\omega_{eff}$ and regarding that $2\Omega_Hr_0=\mathcal{E}$ we find $\omega \sim m\Omega_H$ is acceptable as long as
\begin{eqnarray}
    \mathcal{E}\gg \frac{\zeta}{G_NM^2}\,,
\end{eqnarray}
which is almost always the case even for ``slowly" rotating Kerr-Newman blac holes. For example, taking $GM^2=10^{90}$, while $\zeta=10^{10}$ and $\mathcal{E}=10^{-2}$. The crucial point is now $\omega r_0\approx m\Omega_Hr_0$ is {\it no longer} small. Yet, $\delta \omega\ll 1$ is still acceptable, as this is equivalent to
\begin{equation}
    \delta \omega \lesssim 4\pi T_H ma_0=\frac{mk}{(G_NM^2)(1+k^2)}\sqrt{8\zeta}\ll 1\,,
\end{equation}
implying that the our matching method to attain the analytical expression of greybody factor remains valid.

Let us now return to the conditions for using the effective field theory in the throat and its potential breakdown in the form of imaginary conformal dimension. 
We consider a given channel within a not-so-large energy $E_i$, for large energies we simply apply the semiclassical approximation. The effective angular number and conformal dimension are 
\begin{eqnarray}
    \tilde{l}=\sqrt{\frac{1}{4}+K_{lm}-2\omega^2(3r_0^2+a_0^2)}-\frac{1}{2}\equiv \Delta-1\,,
\end{eqnarray}
while $\omega\in[0,m\Omega_H+E_i]\sim[0,m\Omega_H]$. According to~\cite{Berti:2005gp}, the constant $K_{lm}$ can be expanded as a series in the small $a_0\omega$  regiment, leading to 
\begin{eqnarray}
    K_{lm}=l(l+1)+[h(l+1,m)-h(l,m)](a_0\omega)^2+O((a_0\omega)^4)\,,
\end{eqnarray}
where $h(l,m)$ is 
\begin{eqnarray}
    h(l)=\frac{(l^2-m^2)l}{(2l-1)(l+1/2)}\,.
\end{eqnarray}
By using the approximation conditions introduced above, when $m=0$ we always have the conclusion $\tilde{l}\approx l$ and the theory is rational. When $m\neq 0$, we expect the approximation to hold if the spin-mass ratio is small,  $k\ll 1$. For $k\sim 1$ a separate analysis is required.

After this clarifications of regimes, we can now derive  the  analytical expression for the greybody factor through the matching method. At the asymptotic infinity we have the approximation
\begin{equation}\label{000}
    x^2R_0''+2xR_0'+(p_{-1}x^2+p_1-p_4)R_0=0,
\end{equation}
which gives the analytical solution in terms of Bessel functions
\begin{equation}
\begin{aligned}
    R_0=c_1\mathrm{J}_{\tilde{l}}(\sqrt{p_{-1}}x)+c_2\mathrm{Y}_{\tilde{l}}(\sqrt{p_{-1}}x)\qquad\qquad \qquad \qquad \\\rightarrow -\frac{1}{2x\sqrt{p_{-1}}}[(ic_1+c_2)e^{ix\sqrt{p_{-1}}-\frac{i}{2}\tilde{l}\pi}+(-ic_1+c_2)e^{-ix\sqrt{p_{-1}}+\frac{i}{2}\tilde{l}\pi}]\,,\quad x\rightarrow \infty\,.
\end{aligned}
\end{equation}
Note that in Eq.(\ref{000}) we omit the linear term in $x$ because it only contributes a pure imaginary phaseb in $r$ at infinity. Expanding this expression in the overlapping region we have the following relation
\begin{equation}\label{corela}
c_1=b_1\frac{2^{\tilde{l}+1}\Gamma(\tilde{l}+\frac{3}{2})}{\sqrt{\pi}(\delta\omega)^{\tilde{l}}}-b_2\frac{\sqrt{\pi}(\delta \omega)^{\tilde{l}+1}}{2^{\tilde{l}}\Gamma(\tilde{l}+\frac{1}{2})}\tan \tilde{l}\pi \,,\qquad c_2=-b_2\frac{\sqrt{\pi}(\delta\omega)^{\tilde{l}+1}}{2^{\tilde{l}}\Gamma(\tilde{l}+\frac{1}{2})}\,.
\end{equation}

Following the definition of greybody factor 
\begin{equation}
    \sigma=1-\mathbf{R}=1-\frac{|c_1-ic_2|^2}{|c_1+ic_2|^2}=\frac{-4\mathrm{Im}(c_2c_1^*)}{|c_1|^2+|c_2|^2-2\mathrm{Im}(c_2c_1^*)}\,,
\end{equation}
regarding that we work in the limit $\delta\omega\rightarrow 0$, we can neglect the second term in the expression of $c_1$, thus we need to compute the terms as follows:
\begin{equation}
    b_2b_1^*=\frac{\Gamma(1\pm 2i\sqrt{p_3})\Gamma(\pm 2\tilde{\beta})}{\Gamma(z_1)\Gamma(1-z_1)\Gamma(z_2)\Gamma(1-z_2)}\,,
\end{equation}
where $z_{1,2}=\frac{1}{2}-i\sqrt{p_3}\pm i\sqrt{p_1-p_2+p_3}-\tilde{\beta}$.  Using the identities $\Gamma(z)\Gamma(1-z)=\frac{\pi}{\sin(\pi z)}\,, \,\Gamma(\pm z)=-\frac{\pi}{z\sin(\pi z)}$, we obtain the numerator
\begin{equation}
    \mathrm{Im}(c_2c_1^*)=-\mathrm{Im}(b_2b_1^*)\frac{2\Gamma(1+\tilde{\beta})\delta\omega}{\Gamma(\tilde{\beta})}=-\delta\omega\sqrt{p_3}\,.
\end{equation}
If we restrict to the $m=0$ channel, we can further make the approximation that $p_1-p_2+p_3\approx p_3$ and $\tilde{l}=l$, so there is
\begin{equation}
|c_1|^2+|c_2|^2=\Gamma(1\pm2i\sqrt{p_3})\Big{[}\frac{\Gamma^2(2\tilde{\beta})2^{1+2\tilde{\beta}}\Gamma^{2}(1+\tilde{\beta})(\delta \omega)^{1-2\tilde{\beta}}}{\pi \Gamma^2(\frac{1}{2}+\tilde{\beta})\Gamma(\frac{1}{2}+\tilde{\beta}\pm 2i\sqrt{p_3})}+\frac{\pi \Gamma^2(-2\tilde{\beta})2^{1-2\tilde{\beta}}(\delta\omega)^{1+2\tilde{\beta}}}{\Gamma^2(\frac{1}{2}-\tilde{\beta})\Gamma^2(\tilde{\beta})\Gamma(\frac{1}{2}-\tilde{\beta}\pm 2i\sqrt{p_3})}\Big{]}\,.
\end{equation}
Now consider the relation
$p_3=\frac{\omega_{eff}}{4\pi T_{H}}$, and the limit that $\omega\rightarrow 0$ while $\beta \omega $ finite, we have
\begin{equation}
    \sigma_{l0}\approx \frac{\omega_{\rm eff}}{2T_{H}}\frac{((r_{+}-r_{-})\omega)^{2l+1}}{2^{2l+1}}\Big{|}\frac{\Gamma(l+1)\Gamma(l+1+i\frac{\omega_{\rm eff}}{2\pi T_{H}})}{\Gamma(l+\frac{3}{2})\Gamma(2l+1)\Gamma(1+i\frac{\omega_{\rm eff}}{2\pi T_{H}})}\Big{|}^2\,,
\end{equation}
which precisely coincides with previous results in the literature~\cite{Bonelli:2021uvf,Brown:2024ajk,Maldacena_1997}. 

However, while the above result is rather accurate for $m=0$, as we have illustrated before, for the  $m\neq 0$ channels, we evaluate  $p_1-p_2+p_3\approx (2r_0\omega-\frac{\omega_{eff}}{4\pi T_H})^2\neq p_3$, and $\tilde{l}\neq l$. Then, in the  near superradiance region the greybody factor would be
\begin{equation}\label{sigma scalar}
    \sigma_{lm}\approx \frac{\omega_{\rm eff}}{2T_{H}}\frac{((r_{+}-r_{-})\omega)^{2\tilde{l}+1}}{2^{2\tilde{l}+1}}\Big{|}\frac{\Gamma(\tilde{l}+1+2ir_0m\Omega_H)\Gamma(\tilde{l}+1+i\frac{\omega_{eff}}{2\pi T_{H}}-2ir_0m\Omega_H)}{\Gamma(\tilde{l}+\frac{3}{2})\Gamma(2\tilde{l}+1)\Gamma(1+i\frac{\omega_{eff}}{2\pi T_{H}})}\Big{|}^2\,, 
\end{equation}
here we have replaced $\omega$ in the arguments by $m\Omega_H$ as a good approximation compared to $\frac{\omega_{\rm eff}}{4\pi T_H}$ when $\omega_{\rm eff}\in [0,E_i]$. This property leads to the result that quantum corrected emission rate cannot \emph{always} return to the semiclassical expression near small local energy $\omega_{\rm eff}$, unless we set the condition that
\begin{equation}\label{cen2}
\frac{q_i^2-q_f^2}{2K}= 4\pi T_H r_0 m\Omega_H\,.
\end{equation}
We will prove this can really lead to the correct answer in the following computation, and we note this is exactly compatible with the saddle point approximation we make for the semiclassical results in the microcanonical ensemble. Indeed, with the interpretation that $q_i-q_f=m$, together with the effective temperature given by $T_H=\sqrt{(E_i-q_i^2/2K)E_{brk}/(2\pi^2)}$ Eq.(\ref{cen2}) is equivalent to
\begin{eqnarray}\label{Eq:charges-initial-final}
    q_i\approx q_f\approx \frac{\sqrt{8\zeta}\mathcal{E}}{\sqrt{1+2\mathcal{E}^2}}\,,
\end{eqnarray}
which is nothing but the saddle point solution we found in Eq.(\ref{peak}). 

In these channels we should also include the contribution from the superradiance region to the total emission energy, where $\omega_{eff}<0$, reflection $\mathbf{R}>1$, leading to a negative greybody factor, $\sigma<0$. In general, the semiclassical emission rate is given by
\begin{equation}
\Big{(}\frac{dE}{dtd\omega}\Big{)}\Big{|}_{\mathrm{semi},lm}=\frac{\sigma_{lm}}{2\pi}\cdot\frac{\omega}{e^{\beta\omega_{\rm eff}}-1}\,.
\end{equation}
We now discuss the leading channel for emission. We note the presence of the factor $(\delta\omega)^{2\tilde{l}+1}$ in the greybody contribution. As $\delta\omega\ll1$, in the $m=0$ channel the effective theory is always rational and the main contribution comes from $l=\tilde{l}$ as small as possible. When $m\neq 0$ but $k=a_0/r_0\ll 1$, the conclusion above still makes sense, and the dominant channel would be $m$ as large as possible while $l$ as small as possible. Apart from the influence of conformal dimension, we also need to take the superradiance into account, and this will lead to even larger total energy emitted for larger values of $m$. 

After obtaining the expression for the greybody factor, we are able to determine the coefficient  $\mathcal{N}_{lm}$ before the total expression in Eq.(\ref{05}). If $m=0$, it is convenient to simply consider the bosonic factor absorbing the same $U(1)$ charge contribution to the energy
\begin{equation}\label{par}
\rho(E_f,q_f)|\bra{E_f,q_f}O\ket{E_i,q_i}|^2=\frac{C}{\pi^2}\sinh(2\pi\sqrt{2C(E_i-\omega)})\frac{\Gamma(\Delta\pm i\sqrt{2C(E_i-\omega)}\pm i\sqrt{2CE_i})}{(2C)^{2\Delta}\Gamma(2\Delta)},
\end{equation}
in the semiclassical limit, i.e., $CE=\frac{E}{E_{brk}}\rightarrow \infty$. Now focus on $l=m=0$ case, in which $\Delta=\frac{1}{2}+\tilde{\beta}\approx 1$:
\begin{equation}
\begin{aligned}
    \mathrm{RHS}=\frac{1}{4\pi^2C}\sinh(2\pi\sqrt{2C(E_i-\omega)})\frac{\pi(\sqrt{2C(E_i-\omega)}+\sqrt{2CE_i})}{\sinh(\pi(\sqrt{2C(E_i-\omega)}+\sqrt{2CE_i}))}\frac{\pi(\sqrt{2C(E_i-\omega)}-\sqrt{2CE_i})}{\sinh(\pi(\sqrt{2C(E_i-\omega)}-\sqrt{2CE_i}))}\\
=\omega\frac{\sinh(2\pi\sqrt{2C(E_i-\omega)})}{\cosh(2\pi\sqrt{2CE_i})-\cosh(2\pi\sqrt{2C(E_i-\omega)})}\approx \frac{\omega}{e^{\beta\omega}-1}\,,\qquad \qquad \qquad \qquad
\end{aligned} 
\end{equation}
where we have taken the ETH effective temperature $\beta=\frac{1}{T_H}=\sqrt{\frac{2\pi^2}{E_{brk}E_i}}$. Let us now consider any $l>m=0$, 
\begin{equation}
    \mathrm{RHS}\approx \frac{\omega}{(2C)^{2\Delta-2}\Gamma(2\Delta)}\frac{\sinh(2\pi\sqrt{2C(E_i-\omega)})}{\cosh(2\pi\sqrt{2CE_i})-\cosh(2\pi\sqrt{2C(E_i-\omega)})}\prod_{j=1}^{\Delta-1}(j^2+u_1^2)(j^2+u_2^2)\,, 
\end{equation}
where $u_{1,2}=\sqrt{2C(E_i-\omega)}\pm\sqrt{2CE_i}$, and we approximate $\Delta$ by its closest integer. Then we take the semiclassical limit and find
\begin{equation}
\mathrm{RHS}\approx \frac{\omega}{(2C)^{2\Delta-2}\Gamma(2\Delta)} \frac{(8CE_i)^{\Delta-1}}{e^{\beta\omega}-1}\prod_{j=1}^{\Delta-1}(j^2+(\frac{\beta\omega}{2\pi})^2)\,.
\end{equation}
Compare this result with the fact that
\begin{equation}
\sigma_{l0}=C\pi\omega^{2\Delta}\Big{(}\frac{G_N}{r_0}\Big{)}^{2\Delta-1}(2CE_i)^{\Delta-1}\Big{|}\frac{\Gamma(\Delta)}{\Gamma(\Delta+\frac{1}{2})\Gamma(2\Delta-1)}\Big{|}^2\prod_{j=1}^{\Delta-1}(j^2+(\frac{\beta\omega}{2\pi})^2)\,,
\end{equation}
we find that if we take 
\begin{equation}
    \mathcal{N}_{l0}^2=\frac{\Gamma(2\Delta)}{4\pi}(r_0^2+a_0^2)^{2\Delta-1}\Big{|}\frac{\Gamma(\Delta)}{\Gamma(\Delta+\frac{1}{2})\Gamma(2\Delta-1)}\Big{|}^2\,,
\end{equation}
the quantum emission rate
\begin{equation}
    \frac{d E}{dtd\omega}=2\pi\mathcal{N}_{l0}^2\omega^{2\Delta}\rho(E_i-\omega,q_i)|\bra{E_i-\omega,q_i}\mathcal{O}_{\Delta}\ket{E_i,q_i}|^2\,,
\end{equation}
can precisely reproduce the semiclassical limit in every $l>m=0$ channel. The normalization factor now merely relies on the quantum channel and is unaffected by the Schwarzian quantum number and frequency, which is reasonable given the definition of this constant factor. As we will show in the next subsection, the $s>0$ case is rather similar to that of scalar.

\begin{figure}[htbp]
	\centering
\includegraphics[width=0.48\textwidth]{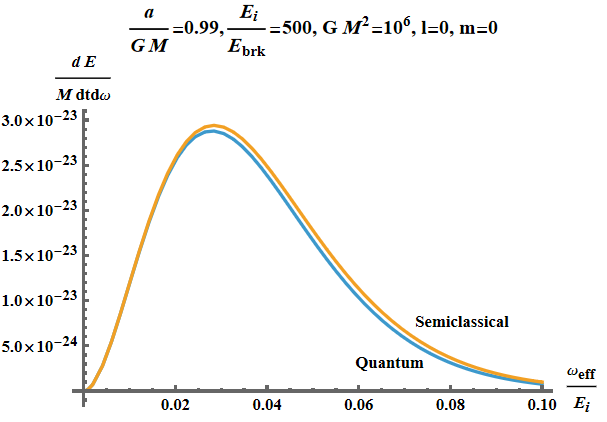}
\includegraphics[width=0.48\textwidth]{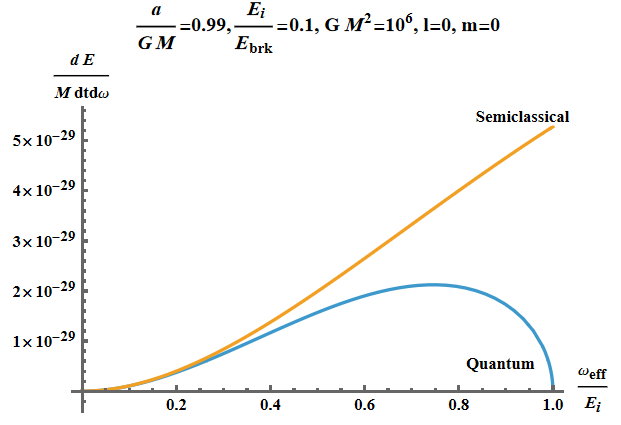}
\caption{Emission rate for the channel $l=m=0$.}
\label{fig:01}
\end{figure}

\begin{figure}[htbp]
	\centering
\includegraphics[width=0.48\textwidth]{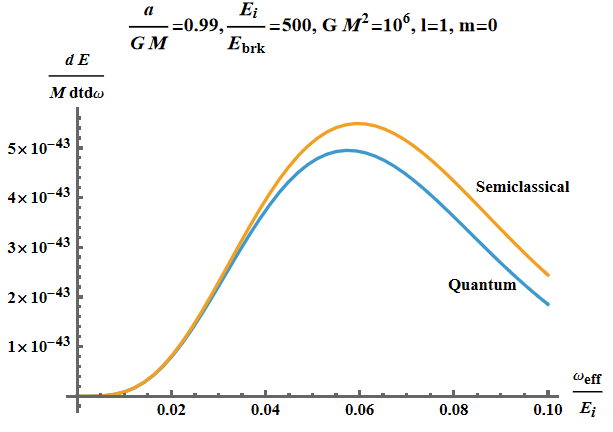}
\includegraphics[width=0.48\textwidth]{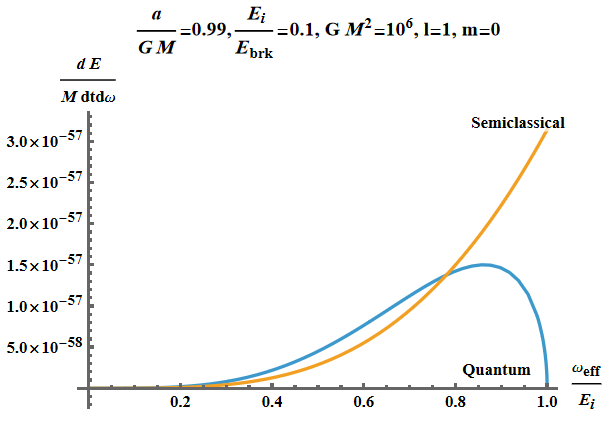}
\caption{Emission rate for the channel $l=1, m=0$.}
\label{fig:02}
\end{figure}

\begin{figure}[htbp]
	\centering
\includegraphics[width=0.49\textwidth]{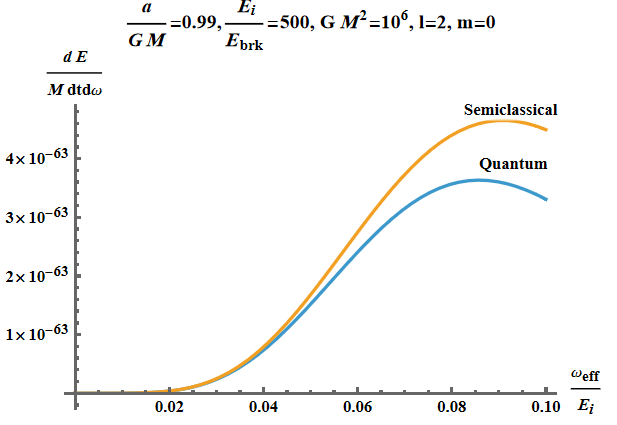}
\includegraphics[width=0.49\textwidth]{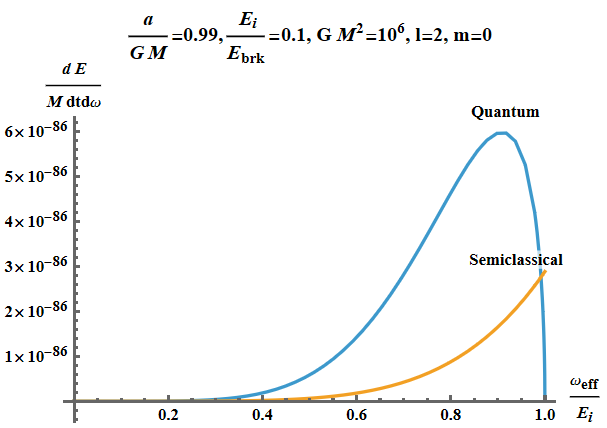}
\caption{Emission rate for the channel $l=2,m=0$.}
\label{fig:03}
\end{figure}

We present plots for scalar emission in various channels. First, we consider the dominant channel  $l=m=0$, as shown in Fig.~\ref{fig:01}. We also present two other sub-dominant channels in Fig.~\ref{fig:02} and Fig.~\ref{fig:03}. In all the plots, the left panels correspond to the semiclassical regime while right panels correspond to quantum regime (in which $E_i/E_{brk}\rightarrow 0$). By comparing the quantum results as shown in the right panels of these three figures, we verify  Page's semiclassical argument that the main contribution comes from the $l=0$ channel in the case where quantum corrections are included. This is not surprising since there is also a factor of $\omega^{2\tilde{l}+1}$ in the coupled Hamiltonian we construct.

For $l>m>0$, the situation is more nuanced. If we make the transformation $E_{i,f}-\frac{q_i^2}{2K}\rightarrow E_{i,f}$, with the condition given by Eq.(\ref{cen2}) we have
\begin{equation}\label{factor note}
\mathrm{RHS}\approx \frac{\tilde{\omega}_{\rm eff}}{(2C)^{2\Delta-2}\Gamma(2\Delta)} \frac{(8CE_i)^{\Delta-1}}{e^{\beta\tilde{\omega}_{\rm eff}}-1}\prod_{j=1}^{\Delta-1}(j^2+(\frac{\beta\tilde{\omega}_{\rm eff}}{2\pi })^2)\,,
\end{equation}
where $\tilde{\omega}_{\rm eff}=\omega_{eff}-4\pi T_H r_0m\Omega_H$.
On the other hand, this time we have
\begin{equation}
\sigma_{lm}=C\pi\omega^{2\Delta-1}\Big{(}\frac{G_N}{r_0}\Big{)}^{2\Delta-1}(2CE_i)^{\Delta-1}\Big{|}\frac{\Gamma(\Delta+2ir_0m\Omega_H)}{\Gamma(\Delta+\frac{1}{2})\Gamma(2\Delta-1)}\Big{|}^2\prod_{j=1}^{\Delta-1}(j^2+(\frac{\beta\tilde{\omega}_{eff}}{2\pi})^2)\tilde{\omega}_{eff}\frac{\sinh\frac{\omega_{\rm eff}}{2T_H}}{\sinh\frac{\tilde{\omega}_{\rm eff}}{2T_H}}\,,
\end{equation}
in which the last factor can be re-expressed as $e^{-2\pi r_0m\Omega_H}(e^{\beta \omega_{\rm eff}}-1)/(e^{\beta \tilde{\omega}_{\rm eff}}-1)$.
Hence if we take 
\begin{equation}\label{normal}
\mathcal{N}_{lm}^2=\frac{\Gamma(2\Delta)}{4\pi}(r_0^2+a_0^2)^{2\Delta-1}\Big{|}\frac{\Gamma(\Delta+2ir_0m\Omega_H)}{\Gamma(\Delta+\frac{1}{2})\Gamma(2\Delta-1)}\Big{|}^2e^{-2\pi r_0m\Omega_H}\,,
\end{equation}
the quantum corrected result
\begin{eqnarray}\label{E infty}
    \frac{dE}{dtd\omega}=2\pi\omega^{2\Delta}\mathcal{N}_{lm}^2\rho(E_i-\omega_{\rm eff},q_i)|\bra{E_i-\omega_{\rm eff},q_i-m}\mathcal{O}_{\Delta}\ket{E_i,q_i}|^2\,.
\end{eqnarray}
The above result can precisely recover the semiclassical result within the full range of Kerr-Newman black holes, from $a_0\approx 0$ to $a_0\approx r_0$. Because now $\Delta$ may change a lot in the whole frequency domain, we take $\Delta$ in Eq.(\ref{normal}) as $\Delta(\tilde{\omega}_{\rm eff}=0)$, in order to precisely  recover the semiclassical expression. Then the normalization factor again is fully determined by the scalar emission channel, as expected.  Finally, we note that the factor $\beta\tilde\omega_{\rm eff}$ in Eq.(\ref{factor note}) can be expressed as $\beta\omega_{\rm eff}-2\pi\mathcal{E}m$ and is in accordance to the exponent $2\pi \mathcal{E}q-\beta E$ we meet in the derivation of the density of states from partition function as in Eq.(\ref{density par}). 

\begin{figure}[htbp]
	\centering
\includegraphics[width=0.49\textwidth]{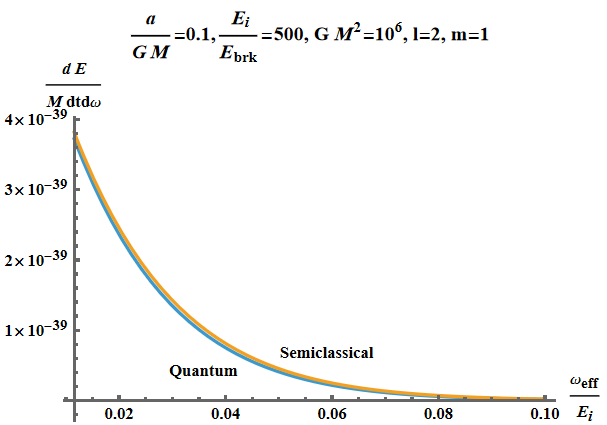}
\includegraphics[width=0.45\textwidth]{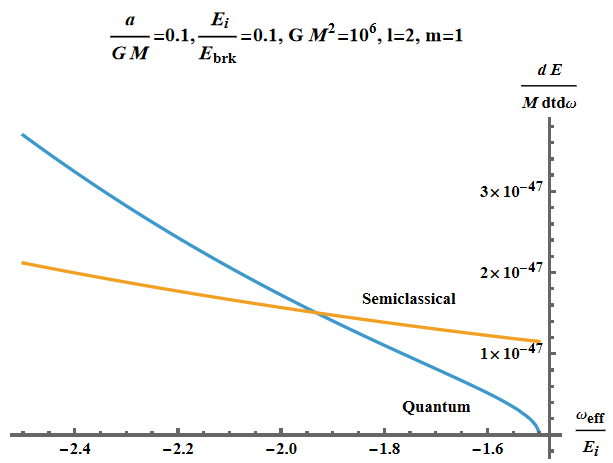}
\caption{Emission rate for the channel $l=2,m=1$.}
\label{fig:04}
\end{figure}

\begin{figure}[htbp]
	\centering
\includegraphics[width=0.45\textwidth]{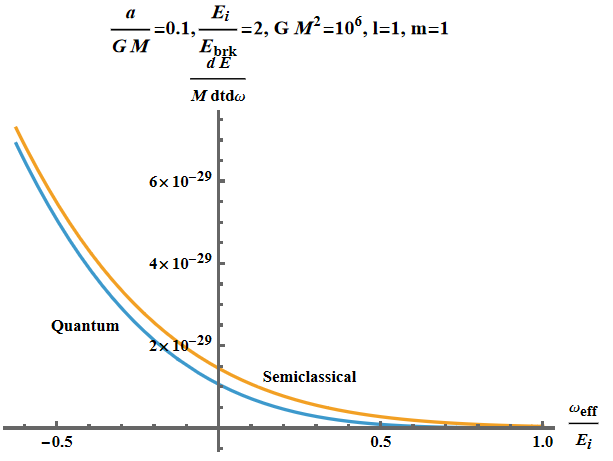}
\includegraphics[width=0.45\textwidth]{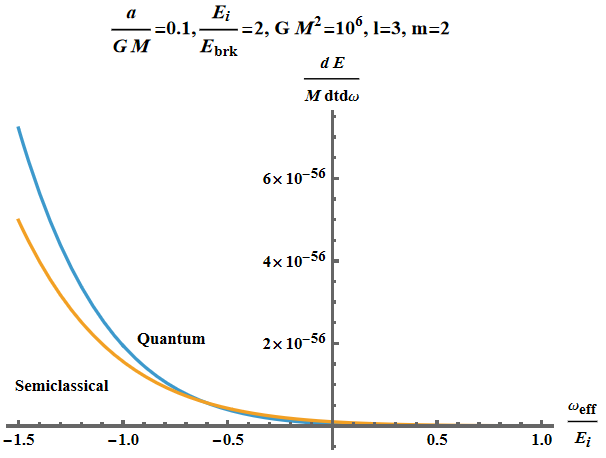}
\caption{Emission rate for the channel $l=1,m=1$ (left) and channel $l=3,m=2$ (right) with $\frac{E_i}{E_{brk}}=2$.}
\label{fig:05}
\end{figure}

We show results for $m\neq 0$ in Fig.~\ref{fig:04} and Fig.~\ref{fig:05}. We only present the $a_0=0.1\,r_0$ case to be able to reliably neglect the influence of rotation in the separation constant and, therefore, in the effective conformal dimension. The left panel of Fig.~\ref{fig:04} is the semiclassical regime of energies as usual, where we have already taken $q_i$ to be the correct expectation value. In this regime the quantum and semiclassical  results are almost indistinguishable. In the right  panel of Fig.~\ref{fig:04} we present the the quantum regime. There are various important points that should be highlighted: (i) note the overall suppression by about eight orders of magnitude, (ii) The range of $\frac{\omega_{\rm eff}}{E_i}$ is negative which is relevant for superradiance and (iii) the quantum result vanishes as expected for a full quantum treatment that prevents the black hole from emitting more energy that it has stored.

In Fig.~\ref{fig:05} we present cases in which $E_i/E_{brk}$ is an $O(1)$ number, in this regime we expect a breakdown of the semiclassical description which also implies that we cannot find a simple relation for the expectation value of $q_{i,f}$ since the fluctuations become more intense. Therefore, we sum over the contribution coming from all possible values of $|q_i|<\sqrt{2KE_i}$ given a fixed $E_i$. There is still a question, about what will be the dominant channel among these superradiant ones and $s$-wave channel $l=m=0$, and we will leave this to Sec.\ref{Subsec:evaporation} where we will provide more details. For now, it is evident from the plots that these channels behave semiclassically when $E_i$ is large enough while the $m<l$ channel is highly suppressed compared to $l=m$. In the given case, with all other parameters being equal, the $l=3, m=2$ channel $(l>m)$ is suppressed by  more than twenty orders of magnitude with respect to the $l=m=1$ one $(l=m)$.

Generally speaking, in the discussion above we have completed the calculation of quantum emission rate for neutral massless scalar fields. For charged ones the steps are rather similar, from the expression in Eq.(\ref{teukolsky}), we see that it suffices to make the replacement
\begin{eqnarray}
    \omega_{eff}=\omega-m\Omega_H\rightarrow \omega-m\Omega_H-e\Phi\,,\qquad 2r_0\omega\rightarrow 2r_0\omega-eQ\,,
\end{eqnarray}
in the semiclassical greybody factor calculated before. On the quantum side we need to make the replacement
\begin{eqnarray}
    E_i-\frac{q_i^2}{2K}\rightarrow E_i-\frac{q_{i,e}^2}{2K_e}-\frac{q_{i,r}^2}{2K_r}\,,
\end{eqnarray}
leading to the effective conformal dimension
\begin{eqnarray}
    \tilde{\beta}=\sqrt{\frac{1}{4}+K_{lm}-2\omega^2(3r_0^2+a_0^2)-e^2Q_0^2+6eQ_0r_0\omega}\,.
\end{eqnarray}
The exchange of charge/angular momentum condition analogous to Eq.(\ref{cen2}) takes the form 
\begin{eqnarray}
    \frac{q_{i,e}^2-q_{f,e}^2}{2K_e}+\frac{q_{i,r}^2-q_{f,r}^2}{2K_r}=2\pi T_H(2r_0m\Omega_H-eQ_0)\,,
\end{eqnarray}
while $T_H$ is now given by $T_H=\sqrt{(E_i-q_{i,e}^2/2K_e-q_{i,r}^2/2K_r)/(2\pi^2)}$, and we take $q_{i,r}-q_{f,r}=m,\,q_{i,e}-q_{f,e}=-e$. We find that the result for the expectation value of the charges before emission exactly satisfies the extremal point condition just like in Eq.(\ref{saddle}) except for one difference that $Q_0\neq \mathcal{E}_e$, but they can be considered approximately equal for small enough rotational parameter, $a_0$. Under this condition, we obtain the expression for the normalization factor as
\begin{equation}\label{normal}
\mathcal{N}_{lm}^2=\frac{\Gamma(2\Delta)}{4\pi}(r_0^2+a_0^2)^{2\Delta-1}\Big{|}\frac{\Gamma(\Delta+2ir_0m\Omega_H-ieQ_0)}{\Gamma(\Delta+\frac{1}{2})\Gamma(2\Delta-1)}\Big{|}^2e^{-\pi(2 r_0m\Omega_H-eQ_0)}\,.
\end{equation}

\subsection{Emission rate of higher spin fields: photon, graviton and spinor}\label{hi}

In the absence of separation of variables for the photon and graviton perturbation in the Kerr-Newman spacetime, we only consider the simpler Kerr case. We now present Teukolsky Master equation for a certain spin-$s$ field  radial function in the coordinate $x=\frac{r-r_{+}}{r_{+}-r_{-}}=(r-r_{+})/\delta$:
\begin{equation}\label{hseq}
    x(x+1)\frac{d^2 R_s}{dx^2}+(s+1)(2x+1)\frac{d R_s}{d x}+\Big{(}\frac{V(x)}{x(x+1)\delta^2}+2am\omega-K_{lm}+4is\omega (r_{+}+\delta x)\Big{)}R_s=0\,,
\end{equation}
with
\begin{equation}
    V(x)=K(x)^2-2is(r_{+}+\delta x-M)K(x)\,,
\end{equation}
\begin{equation}
    K(x)=[(r_{+}+\delta x)^2+a^2]\omega-am\,,
\end{equation}
while the angular part is given by ($y=\cos\theta$)
\begin{eqnarray}
    \frac{d}{dy}((1-y^2)\frac{d S_s}{dy})+\Big{(}s-2a\omega sy-a^2\omega^2(1-y^2)-\frac{(m+sy)^2}{1-y^2}+K_{lm}\Big{)}S_s=0\,.
\end{eqnarray}

By the coordinate transform $\rho_s=\Delta^{\frac{s}{2}}R_s$ (note that here $\Delta=(r-r_{+})(r-r_{-})$ and should not be confused with the effective conformal dimension introduce before), we have the following equation for the radial function
\begin{equation}
    x(x+1)\frac{d^2\rho_s}{dx^2}+(2x+1)\frac{d\rho_s}{dx}+\Big{(}\frac{p_{-1}x^4+p_0x^3+p_1x^2+p_2x+p_3}{x(x+1)}-p_4\Big{)}\rho_s=0\,,
\end{equation}
in which
\begin{equation}
    p_{-1}=\delta^2\omega^2\,, \qquad p_0=4\delta\omega^2r_{+}+2is\delta\omega\,,
\end{equation}
\begin{equation}
    p_1=-s^2-4isr_{+}\omega+3is\delta\omega+2\omega[(3r_{+}^2+a^2)\omega-am]\,,
\end{equation}
\begin{equation}
    p_2=\frac{2\omega r_{+}-is}{\delta}(2(r_{+}^2+a^2)\omega-2am-i\delta s)\,,
\end{equation}
\begin{equation}
    p_3=\frac{(2am+i\delta s-2(a^2+r_{+}^2)\omega)^2}{4\delta^2}\,,
\end{equation}
\begin{equation}
    p_4=K_{lm}-2am\omega+s-4isr_{+}\omega\,.
\end{equation}
As a result of the above relations, one has 
\begin{equation}
p_1-p_2+p_3=(\frac{\omega_{eff}}{4\pi T_H}-\frac{is}{2})^2+2(is-2r_0\omega)(\frac{\omega_{eff}}{4\pi T_H}-r_0\omega)=(\frac{\omega_{eff}}{4\pi T_H}+\frac{is}{2}-2r_0\omega)^2\,.
\end{equation}
We have the same asymptotic form of $\rho_s$ as in the scalar case, but now the conformal dimension is altered. Noting that now for Kerr black holes,  $r_0=a_0$, we obtain
\begin{equation}\label{cdf}
    \tilde{l}\approx -\frac{1}{2}+\tilde{\beta}=\sqrt{\frac{1}{4}+K_{lm}+s(s+1)-8\omega^2r_0^2}-\frac{1}{2}\neq l\,,
\end{equation}
in which $8\omega^2r_0^2\lesssim 2m^2$ for not-so-large $E_i$. The separation constant,  $K_{lm}$, can be expressed as
\begin{eqnarray}\label{ana dim}
    K_{lm}=l(l+1)-s(s+1)+[h(l+1,m,s)-h(l,m,s)](r_0\omega)^2+O(r_0\omega)^4\,,
\end{eqnarray}
in the  small $\omega$ limit, in which $h(l,m,s)$ is a certain function presented in~\cite{Berti:2005gp}. The dominant channel would still be $l\geq s$ as small as possible with $m < l$ as large as possible.

A recent outstanding development in applications of certain mathematical physics results allows to reinterpret and solve Teukolsky's master equation  \cite{Bonelli:2021uvf,Arnaudo:2024bbd,Arnaudo:2024rhv}. With the same definition of the variable $x$ as in Eq.(\ref{hseq}), we instead choose to set $\chi(x)=\Delta^{\frac{s+1}{2}}R_s(r)$. Then we have the Heun equation 
\begin{eqnarray}
    \chi''(x)+V_r(x)\chi(x)=0\,,\qquad V_r(x)=\frac{1}{x^2(x-1)^2}\sum_{i=0}^{4}A_ix^i=0\,,
\end{eqnarray}
in which $A_i$ are those related to the  Belavin-Polyakov-Zamolodchikov equation in conformal field theory 
\begin{align*}
   & A_0=\frac{1}{4}-a_1^2\,,\qquad
    A_1=-\frac{1}{4}+E+a_1^2-a_2^2-m_3\Lambda\,,\\
    &A_2=\frac{1}{4}-E+2m_3\Lambda-\frac{\Lambda^2}{4}\,,\quad
A_3=-m_3\Lambda+\frac{\Lambda^2}{2}\,,\quad
    A_4=-\frac{\Lambda^2}{4}\,.
\end{align*}
and we further have the dictionary between CFT and gravity theory as
\begin{align*}
   & E=\frac{1}{4}+K_{lm}+s(s+1)-8M^2\omega^2-(2M\omega^2+is\omega)\delta\,,\\
   & a_1=-i\frac{\omega_{eff}}{4\pi T_H}+2iM\omega+\frac{s}{2}\,,\qquad a_2=-i\frac{\omega_{eff}}{4\pi T_H}-\frac{s}{2}\,,\\
   & m_3=-2iM\omega+s\,,\qquad \Lambda=-2i\delta\omega\,.
\end{align*}
Similar identifications are established for the angular function $S_{lm}(\theta,\varphi)$.

The general solution presented in Eq.(\ref{general}), in the near-horizon region,takes the following form:
\begin{equation}\label{near horizon}
R_s\rightarrow \delta^{s+i\frac{\omega_{eff}}{4\pi T_H}}\Delta^{-s}(r-r_{+})^{-i\frac{\omega_{eff}}{4\pi T_H}}\,.
\end{equation}
In the overlapping region, it becomes 
\begin{equation}\label{barrier}
\begin{aligned}
\rho_s \rightarrow 
\frac{\Gamma(2\tilde{\beta})\Gamma(1-s+i\frac{\omega_{eff}}{2\pi T_H})}{\Gamma(\frac{1}{2}+\tilde{\beta}+i\frac{\omega_{eff}}{2\pi T_H}-2ir_0\omega)\Gamma(\frac{1}{2}+\tilde{\beta}-s+2ir_0\omega)}x^{\tilde{l}}\\+\frac{\Gamma(2\tilde{\beta})\Gamma(1-s+i\frac{\omega_{eff}}{2\pi T_H})}{\Gamma(\frac{1}{2}-\tilde{\beta}+i\frac{\omega_{eff}}{2\pi T_H}-2ir_0\omega)\Gamma(\frac{1}{2}-\tilde{\beta}-s+2ir_0\omega)}x^{-\tilde{l}-1}\,.
\end{aligned}
\end{equation}
In the far region the radial equation takes the form 
\begin{equation}\label{eq1}
    x^2\frac{d^2\rho_s}{dx^2}+2x\frac{d\rho_s}{dx}+(p_{-1}x^2+p_0x+p_1-p_4)\rho_s=0\,,
\end{equation}
note that here $p_0$  is a complex number and thus has nontrivial effects in the power dependence of the solution on $r$ at large distance. With $\delta \omega x\rightarrow\infty$ we have the following solution of Eq.(\ref{eq1})
\begin{equation}
\begin{aligned}
    \rho_s=d_+x^{\tilde{l}}e^{-i\sqrt{p_{-1}}x}U[\tilde{l}+1-\frac{ip_0}{2\sqrt{p_{-1}}},2(\tilde{l}+1),2i\sqrt{p_{-1}}x]+d_-x^{\tilde{l}}e^{i\sqrt{p_{-1}}x}U[\tilde{l}+1+\frac{ip_0}{2\sqrt{p_{-1}}},2(\tilde{l}+1),-2i\sqrt{p_{-1}}x]
    \\\rightarrow d_+(\delta \omega)^{-\tilde{l}}(2i)^{-\tilde{l}-1+s}(\omega r)^{s-1}e^{-i\omega r_{*}}+d_-(\delta \omega)^{-\tilde{l}}(-2i)^{-\tilde{l}-1-s}(\omega r)^{-s-1}e^{i\omega r_{*}}\,,\qquad\qquad\qquad
\end{aligned}
\end{equation}
where $U$ are confluent hypergeometric functions and the radial coordinate behaves as $r\sim \delta x$. In the region  $\delta \omega x\rightarrow 0$ we have 
\begin{equation}
    \rho_s\rightarrow b_1x^{\tilde{l}}+b_{2}x^{-\tilde{l}-1}\,,
\end{equation}
in which
\begin{equation}
b_{1}=\frac{\Gamma(-1-2\tilde{l})}{\Gamma(-\tilde{l}-s)}(d_++d_-)\,,\qquad 
b_{2}=(i\delta\omega)^{-1-2\tilde{l}}\frac{2^{-1-2\tilde{l}}\Gamma(1+2\tilde{l})}{\Gamma(1+\tilde{l}-s)}[d_++(-1)^{-1-2\tilde{l}}d_-]\,.
\end{equation}
Note that here we omit the real part of $p_0$ to keep accordance to the scalar case. Inverting the above equations  yields 
\begin{equation}\label{far}
    d_{\pm}=ie^{\frac{i\pi}{2}(2\tilde{l}+1)}\Big{\{}\frac{2^{2\tilde{l}+1}\Gamma(\tilde{l}+\frac{3}{2})}{\sqrt{\pi}\prod_{j=1}^{s-1}(-\tilde{l}-j)}b_1+\frac{\sqrt{\pi}(\mp i\delta\omega)^{2\tilde{l}+1}}{\Gamma(\tilde{l}+\frac{1}{2})\sin[(\tilde{l}+\frac{1}{2})\pi]\prod_{j=1}^{s-1}(1+\tilde{l}-j)}b_2\Big{\}}\,,
\end{equation}
and one is able to verify that in the special case $s=m=0$ this precisely reproduces the coefficient relation presented in Eq.(\ref{corela}).

The final expression for the greybody factor is slightly  more involved. For photon emission we have
\begin{equation}
   \sigma_{lm,s=1}=\frac{\omega}{2GMr_{+}\omega_{eff}}\Big{|}\frac{R_{{abs}}}{R_{in}}\Big{|}^2\,,
\end{equation}
where the coefficients are determined by the asymptotic behavior 
\begin{equation}
\begin{aligned}
     \phi_0 \sim e^{im\varphi-i\omega t}S_{lm}(\theta)(R_{in}\frac{e^{-i\omega r_{*}}}{r}+R_{out}\frac{e^{i\omega r_{*}}}{r^3})\,, \; r\rightarrow \infty\,,\\
     \sim e^{im\varphi -i\omega t}S_{lm}(\theta)R_{abs} \Delta^{-1}(r-r_{+})^{-i\frac{\omega_{eff}}{4\pi T_H}}\,,\; r \rightarrow r_{+}\,.
\end{aligned}
\end{equation}
By combing the results in Eq.(\ref{near horizon}), Eq.(\ref{barrier}) and Eq.(\ref{far}) for $s=1$ we obtain
\begin{equation}\label{sigma photon}
\begin{aligned}
    \sigma_{lm,s=1}=\frac{2\pi^2T_H(\tilde{l}+1)^2}{\omega_{eff}}\frac{((r_{+}-r_{-})\omega)^{2\tilde{l}+1}}{2^{2\tilde{l}+1}}\Big{|}\frac{\Gamma(\tilde{l}+2ir_0m\Omega_H)\Gamma(\tilde{l}+1+i\frac{\omega_{eff}}{2\pi T_H}-2ir_0m\Omega_H)}{\Gamma(\tilde{l}+\frac{3}{2})\Gamma(2\tilde{l}+1)\Gamma(i\frac{\omega_{eff}}{2\pi T_H})}\Big{|}^2\\
    =\frac{(\tilde{l}+1)^2}{\tilde{l}^2+(2r_0m\Omega_H)^2}\frac{\omega_{eff}}{2T_H}\frac{((r_{+}-r_{-})\omega)^{2\tilde{l}+1}}{2^{2\tilde{l}+1}}\Big{|}\frac{\Gamma(\tilde{l}+1+2ir_0m\Omega_H)\Gamma(\tilde{l}+1+i\frac{\omega_{eff}}{2\pi T_H}-2ir_0m\Omega_H)}{\Gamma(\tilde{l}+\frac{3}{2})\Gamma(2\tilde{l}+1)\Gamma(1+i\frac{\omega_{eff}}{2\pi T_H})}\Big{|}^2\,,
\end{aligned}
\end{equation}
and then by comparing Eq.(\ref{sigma photon}) with Eq.(\ref{sigma scalar}) one finds the relation
\begin{equation}
    \sigma_{lm,s=1}=\frac{(\tilde{l}+1)^2}{\tilde{l}^2+m^2}\sigma_{l^{*}m,s=0}\,,
\end{equation}
where we have taken $\Omega_H=1/2r_0$ for extremal Kerr black hole with no electric charge. Here $l^{*}$ is a quantum number such that $\tilde{l}(l^*,s=0)=\tilde{l}(l,s=1)$. It is notable that our result matches Page's work precisely~\cite{Page:1976df,Page:1976ki}. These two greybody factors are just proportional to each other, with the only manifest influence of spin coming in the  conformal dimension. This result is intuitively clear, from the Schwarzian point of view it can be explained  by the fact that the quantum Schwarzian matrix element is only manifestly altered by nonzero spin, $s$, via the conformal dimension. We conclude that the normalization 
\begin{equation}
    \mathcal{N}_{lm,s=1}^2=\frac{(\tilde{l}+1)^2}{\tilde{l}^2+m^2}\mathcal{N}_{l^*m,s=0}^2\,,
\end{equation}
is sufficient to guarantee the quantum emission rate is able to return to its semiclassical limit in the large $E_i$ regime. Note that here we always have $\mathcal{N}_{lm}^2$ to rely on $\tilde{l}$ only, so we can just multiply what we obtained like in Eq.(\ref{normal}) with a constant factor.

Similar results can be obtained for the graviton leading to 
\begin{equation}
    \sigma_{lm,s=2}=\frac{\omega^3}{(2GMr_{+})^3\omega_{eff}(\omega_{eff}^2+4\pi^2T_{H}^2)}\Big{|}\frac{R_{{abs}}}{R_{in}}\Big{|}^2\,,
\end{equation}
where the precise asymptotics requires 
\begin{equation}
\begin{aligned}
     \psi_0 \sim e^{im\varphi-i\omega t}S_{lm}(\theta)(R_{in}\frac{e^{-i\omega r_{*}}}{r}+R_{out}\frac{e^{i\omega r_{*}}}{r^5})\,, \; r\rightarrow \infty\,,\\
     \sim e^{im\varphi -i\omega t}S_{lm}(\theta)R_{abs} \Delta^{-2}(r-r_{+})^{-i\frac{\omega_{eff}}{4\pi T_H}}\,,\; r \rightarrow r_{+}\,.
\end{aligned}    
\end{equation}
By combining the results in Eq.(\ref{near horizon}), Eq.(\ref{barrier}) and Eq.(\ref{far}) for $s=2$ we obtain
\begin{equation}
\begin{aligned}
\sigma_{lm,s=2}=\frac{8\pi^4T_H^3(\tilde{l}+1)^2(\tilde{l}+2)^2}{\omega_{eff}(\omega_{eff}^2+4\pi^2T_H^2)}\frac{((r_+-r_{-})\omega)^{2\tilde{l}+1}}{2^{2\tilde{l}+1}}\Big{|}\frac{\Gamma(\tilde{l}-1+2ir_0m\Omega_H)\Gamma(\tilde{l}+1+i\frac{\omega_{eff}}{2\pi T_H}-2ir_0m\Omega_H)}{\Gamma(\tilde{l}+\frac{3}{2})\Gamma(2\tilde{l}+1)\Gamma(-1+i\frac{\omega_{eff}}{2\pi T_H})}\Big{|}^2\\
=\frac{(\tilde{l}+1)^2}{\tilde{l}^2+(2r_0m\Omega_H)^2}\frac{(\tilde{l}+2)^2}{(\tilde{l}-1)^2+(2r_0m\Omega_H)^2}\sigma_{l^*m,s=0}\,.\qquad\qquad\qquad\qquad
\end{aligned}
\end{equation}
Taking 
\begin{equation}
    \mathcal{N}_{lm,s=2}^2=\frac{(\tilde{l}+1)^2}{\tilde{l}^2+m^2}\frac{(\tilde{l}+2)^2}{(\tilde{l}-1)^2+m^2}\mathcal{N}_{l^*m,s=0}^2\,,
\end{equation}
we are able to precisely reproduce the semiclassical result in large $E_i$ regime.

The discussion above is also applicable to the fermion case, where both angular number $l$ and spin $s$ have to take half-integer values, and it is notable that there is the well-known absence of superradiant effect for fermions. For instance, we derive the result of spinor greybody factor to that similar to Page's expression \cite{Page:1976ki}:
\begin{eqnarray}
    \sigma_{lm,\frac{1}{2}}\approx \frac{\pi}{2}\frac{((r_{+}-r_{-})\omega)^{2\tilde{l}+1}}{2^{2\tilde{l}+1}}\Big{|}\frac{\Gamma(\tilde{l}+\frac{1}{2}+2ir_0m\Omega_{H})\Gamma(\tilde{l}+1+i\frac{\omega_{\rm eff}}{2\pi T_{H}}-2ir_0m\Omega_H)}{\Gamma(\tilde{l}+1)\Gamma(2\tilde{l}+1)\Gamma(\frac{1}{2}+i\frac{\omega_{\rm eff}}{2\pi T_{H}})}\Big{|}^2\,, 
\end{eqnarray}
and the requirement that the quantum corrected result
\begin{eqnarray}\label{E infty}
    \frac{dE}{dtd\omega}=2\pi\omega^{2\Delta}\mathcal{N}_{lm}^2\rho(E_i-\omega_{eff},q_i-m)|\bra{E_i-\omega_{\rm eff},q_i-m}\mathcal{O}_{\Delta}\ket{E_i,q_i}|^2\,,
\end{eqnarray}
recovers the semiclassical expression
\begin{equation}
\Big{(}\frac{dE}{dtd\omega}\Big{)}\Big{|}_{\mathrm{semi},lm}=\frac{\sigma_{lm}}{2\pi}\cdot\frac{\omega}{e^{\beta\omega_{\rm eff}}+1}\,,
\end{equation}
determines the normalization factor to be 
\begin{eqnarray}
    \mathcal{N}_{lm,s=\frac{1}{2}}^2=\frac{\Gamma(2\Delta)}{8\pi}(r_0^2+a_0^2)^{2\Delta-1}\Big{|}\frac{\Gamma(\Delta-\frac{1}{2}+2ir_0m\Omega_H)}{\Gamma(\Delta)\Gamma(2\Delta-1)}\Big{|}^2e^{-2\pi r_0m\Omega_H}\,.
\end{eqnarray}

\subsection{Di-particle emission and pair production}
We now consider the situation in which the initial and final state have energy difference $\omega+\omega'$, but the quantum number $q$ is unchanged. Two adjoint emission processes are responsible for this type of decay, leading to: (i) the emission of a particle with $(\omega,m)$ and then with $(\omega',-m)$, or (ii)  exchange $\omega$ and $\omega'$. Here we only take $m\neq0$ into account as these are dominant channels for $l\geq s>0$. This transition is a second-order process and the transition rate involves two more integrals in energy and two more sums over the $U(1)$ quantum number. The formula is~\cite{Brown:2024ajk}
\begin{eqnarray}\label{trans rate is}
\begin{aligned}
    \Gamma_{i\rightarrow f}=2\pi \Big{|}\sum_{I}\sum_{\omega_I=\omega,\omega'}\frac{\bra{E_f,q_i,\omega,\omega'}H_I\ket{E_I,q_I,\omega_I}\bra{E_I,q_I,\omega_I}H_I\ket{E_i,q_i}}{E_I+\omega_I-E_i}\Big{|}^2\\
    \times\delta(E_f+\omega+\omega'-E_i)\,.
\end{aligned}
\end{eqnarray}
With a $U(1)$ sector composition law one is able to express it as (with fixed $l$)
\begin{eqnarray}\label{gammaif}
\begin{aligned}
    \Gamma_{i\rightarrow f}=2\pi\sum_{m,m'\neq 0}\mathcal{N}_{l,\pm m}\mathcal{N}_{l,\pm m'}\omega^{2\Delta-1}\omega'^{2\Delta-1}\delta(E_f+\omega+\omega'-E_i)\int_{0}^{\infty}dE_I\rho(E_I,q_i-m)\qquad \qquad \\
\times \int_{0}^{\infty}dE_{I'}\rho(E_{I'},q_i-m')\sum_{\{\omega_I,\omega_I'\}}\frac{\mathcal{A}(E_i,E_f,q_i;E_I,E_{I'},m,m')}{(E_I+\omega_{I}-E_i)(E_{I'}+\omega_{I'}-E_i)}\,,\qquad\qquad
\end{aligned}
\end{eqnarray}
where, following standard notation, we denote  $\mathcal{N}_{l\pm m}=\mathcal{N}_{lm}\mathcal{N}_{l,-m}$, and
\begin{eqnarray}
\begin{aligned}
    \mathcal{A}(E_i,E_f,q_i;E_I,E_{I'},m,m')=
     \bra{E_i,q_i}\mathcal{O}\ket{E_{I'},q_i-m'}\bra{E_{I'},q_i-m'}\mathcal{O}\ket{E_f,q_i}\\\times
    \bra{E_f,q_i}\mathcal{O}\ket{E_I,q_i-m}\bra{E_I,q_i-m}\mathcal{O}\ket{E_i,q_i}\,.
\end{aligned}    
\end{eqnarray}
Generally, through the duality and diagrammatic rules for four point functions, can receive contributions from two TOCs (Time-Ordered Correlators) and one OTOC (Out-of-Time-Ordered Correlator) Wick contraction. As we are only interested in the $E_i\neq E_f$ case, we have 
\begin{eqnarray}\label{4pt function}
\begin{aligned}
    \mathcal{A}(E_i,E_f,q_i;E_I,E_{I'},m,m')=\mathcal{N}_{4\mathrm{pt}}E_{brk}^{4\Delta}(\Gamma^{\Delta}_{i,I'}\Gamma^{\Delta}_{I',f}\Gamma^{\Delta}_{f,I}\Gamma^{\Delta}_{I,i})^{1/2}\Big{(}\frac{\delta(E_I-E_{I'})}{e^{-S_0}\rho(E_I,q_I)}+\Big{\{}\Delta,i,f,I,I'\Big{\}}\Big{)}\,,
\end{aligned}    
\end{eqnarray}
in which $\mathcal{N}_{4\mathrm{pt}}=(2^{2\Delta-1}\Gamma(2\Delta))^{-2}e^{-3S_0}$, and
\begin{eqnarray}
    \Gamma^{\Delta}_{i,f}=\Gamma(\Delta\pm i\sqrt{2C(E_i-\frac{q_i^2}{2K})}\pm i\sqrt{2C(E_f-\frac{q_f^2}{2K})})\,,
\end{eqnarray}
(here we have taken $q_{I'}=q_I=q_i-m$), and
\begin{eqnarray}
\begin{aligned}
    \Big{\{}\Delta,i,f,I,I'\Big{\}}=(\Gamma^{\Delta}_{i,I'}\Gamma^{\Delta}_{I',f}\Gamma^{\Delta}_{f,I}\Gamma^{\Delta}_{I,i})^{1/2}\qquad\quad\qquad\qquad\qquad\qquad\\\times\mathcal{W}_{\sqrt{2C(E_{I}-q_{I}^2/2K)}}\Big{(}\sqrt{2C(E_{I'}-q_{I'}^2/2K)},\Delta\pm i\sqrt{2C(E_{i}-q_{i}^2/2K)},\Delta\pm i\sqrt{2C(E_{f}-q_{i}^2/2K)}\Big{)}\,,
\end{aligned}
\end{eqnarray}
and in the above expression all Heaviside theta functions accompanied with the square root are omitted. We can then write out the full emission rate as the above transition rate multiply by all possible final state $E_f$ and then frequency domain $\omega,\omega'$:
\begin{eqnarray}\label{di-particle}
    \frac{dE}{dt}=\int_0^{E_i}dE_f\rho(E_f,q_i)\int_0^{E_i-E_f}d\omega'd\omega(\omega+\omega')\Gamma_{i\rightarrow f}\,.
\end{eqnarray}

We can obtain a significantly simpler expression for the quantum regime, in which $E_i\sim E_f\ll E_{brk}, q_i=0$, so $\omega$ and $\omega'$ is also rather small. For all terms involving not only $E_{i,f}$ but $E_{I}$ as well, we can neglect the former as $E_I\geq \frac{m^2}{4}E_{brk}$. As an example we treat di-photon emission in Kerr spacetime, in which $m,m'$ are summing over $\pm1$, and now $\Delta\sim 2$ because the frequency domain we are interested in has $\omega r_0\rightarrow 0$. Then we can approximate the TOC part of Eq.(\ref{di-particle}) as
\begin{eqnarray}\label{TOC}
\begin{aligned}
    \frac{dE}{dt}\Big{|}_{\mathrm{TOC}}\approx\frac{256r_0^{12}E_{brk}^4}{81\pi\sinh^2\pi}\int_0^{E_i-E_f}d\omega\omega^3(E_i-E_f-\omega)^3\int_0^{E_i}dE_f(E_i-E_f)\sinh(2\pi\sqrt{2CE_f})\qquad\\
    \times \int_0^{\infty}dE_I\frac{E_I^2(1+2CE_I)^4}{(E_I+E_{brk}/4)^2}\frac{\sinh(2\pi\sqrt{2CE_I})}{\sinh^4(\pi\sqrt{2CE_I})}\,,\qquad\qquad\qquad\\
\end{aligned}
\end{eqnarray}
which can be numerically calculated to be $5.29\times10^{-7}r_0^{12}E_{brk}^{9/2}E_i^{19/2}$. At the same time, the OTOC part is
\begin{eqnarray}\label{OTOC}
\begin{aligned}
     \frac{dE}{dt}\Big{|}_{\mathrm{OTOC}}\approx\frac{32\pi r_0^{12}E_{brk}^5}{81\sinh^2\pi} \int_0^{E_i-E_f}d\omega\omega^3(E_i-E_f-\omega)^3\int_0^{E_i}dE_f(E_i-E_f)\sinh(2\pi\sqrt{2CE_f})\qquad\qquad\\
     \times \int_0^{\infty}dE_IdE_{I'}\mathcal{W}_{\sqrt{2CE_I}}(\sqrt{2CE_{I'}},2,2,2,2)\Big{[}\Big{(}\frac{\sqrt{2CE_I}(1+2CE_I)}{\sinh (\pi\sqrt{2CE_I})}\Big{)}^4
     \frac{\sinh(2\pi\sqrt{2CE_I})}{E_I+E_{brk}/4}\times(I\leftrightarrow I')\Big{]}\,,\qquad
\end{aligned}
\end{eqnarray}
and by numerical integral it gives $5.72\times 10^{-7}r_0^{12}E_{brk}^{9/2}E_i^{19/2}$. Similar treatment can be applied to di-graviton and di-spinor emission. 

We can, through the same method, compute the pair production rate of the extremal Kerr-Newman black hole. The process can be described exactly by the di-particle emission with the only alteration coming from replacing angular momentum charge $q_r$ with electric charge $q_e$. We can then compute $d<\mathrm{pair}>/dt$ and compare the results with that of the semiclassical approximation. Let us write it out as
\begin{eqnarray}
    \frac{d<\mathrm{pair}>}{dt}=\int_0^{E_f}dE_f\rho(E_f,q_i)\int_0^{E_i-E_f}d\omega'd\omega\Gamma_{i\rightarrow f}\,.
\end{eqnarray}
where $\Gamma_{i\rightarrow f}$ is similarly derived as in Eq.(\ref{trans rate is}). Now there is no need for considering $l>0$ channel as we only control the leading order contribution. 

We end the discussion of the di-particle emission by commenting on several subtleties. First is, if we take into account higher genus contributions, it seems that according to Saad-Shenker-Stanford duality interpretation~\cite{Saad:2019lba}, in which JT gravity should dual to a double scaled matrix integral whose dimension is proportional to $e^{S_0}$, then it leads to the expectation value of the product of two density of states receiving nontrivial quantum correction. Specifically, we replace (we first absorb the ground energy into $E_I$) $\rho(E_{I'})\rho(E_I)$ in Eq.(\ref{gammaif}) by the ``sine kernel formula"
\begin{eqnarray}
    <\rho(E_I)\rho(E_{I'})>\approx \rho_0(E_I)\rho_0(E_{I'})-\frac{\sin^2(\pi\rho_0(E_I)(E_I-E_{I'}))}{\pi^2(E_I-E_{I'})^2}+\rho_0(E_I)\delta(E_I-E_{I'})\,.
\end{eqnarray}
Generally, the integral pertaining to the TOC part will have singular behavior, although the divergence comes from topologically sub-leading term
$\rho_0(E_I)\delta(E_I-E_{I'})$. Actually this puzzle results from naively adopting the generic formula. If we regard the entirety of the four point function times two density of states as an ensemble average, then eventually no divergence survives because the delta function in the TOC part of Eq.(\ref{4pt function}) already imposes energy conservation and we have to sum over one internal energy, encountering merely one density\footnote{One of the authors thanks Z.Yang for this insight.}. Intuitively, this can be interpreted as a result of the matrix element in energy basis being also affected by geometric fluctuations via the length of geodesics~\cite{Yang:2018gdb}.

Another puzzle arises from selecting intermediate frequencies $\omega_I$ and $\omega_{I'}$ in $\{\omega,\omega'\}$ in Eq.~\eqref{trans rate is}. These frequencies are not actually the local frequencies according to the Frolov-Thorne vacuum, leading to slightly non-standard treatment. We, therefore, face a dilemma:  either we choose $\{\omega,\omega'\}$ to be small and positive but the term in the denominator $E_I+\omega_I-E_i$ will vanish at some $E_I$, and the whole expression requires some regularization; or, we choose $\{\omega_{eff},\omega_{eff}'\}$ to be small positive but then the Hawking quanta with negative $m$ will not arrive at infinity and one does not know how to quantize this field according to the holographic dictionary. In general this reflects the superradiance instability, which leads to more acute contradictions when computing higher loop processes. 

In conclusion, the di-particle computation is potentially problematic. Regardless of these problems coming from using the Schwarzian effective theory to study such a process, we expect a final well-behaved di-particle emission to have higher order dependence on the initial energy $E_i$ from dimensional analysis. For photon this should be $\sim E_i^{\frac{19}{2}}$. In the next section, through comparison to that of single particle emission we will see that such process is generally inessential in the quantum region of parameters, unlike for the case of  Reissner-Nordstrom black holes~\cite{Brown:2024ajk}. 

\section{Evaporation Process of a Near Extremal Rotating Black Hole}\label{Sec:Evaporation}

In this section we use the formulas obtained in Section \ref{Sec:EmissionRates} to track the evaporation history of near-extremal rotating black holes.  In particular, in the last subsection, \ref{Subsec:evaporation}, we present some comparisons with the semi-classical results as well as with the quantum-corrected evaporation results for spherically symmetric black holes.

\subsection{Energy emitted}
We first focus on a special process in which there is no $U(1)$ charge exchange during evaporation. We discussed the energy flux of di-particle emission in Section \ref{Sec:EmissionRates}. Here, we will prove that the dominant contributions still come from all possible $m=0,\,l=s$ channels of single particle emission. This result mimics Page's conclusion from  half a century ago \cite{Page:1976df,Page:1976ki}, indicating the dominant role of the greybody factors even in the quantum corrected framework. We work in the approximation where the total spin and charge of the near-extremal black hole is fixed while the energy is radiated until it arrives to the precise extremal black hole configuration. In such a process we can always eliminate the influence of the choice for $q_i$ through ground energy redefinition. Now our starting point will be a Kerr (Newman) black hole  with $E_i \gtrsim E_{brk}$ as its microcanonical energy. Recall that we parametrize the initial energy as 
\begin{equation}
    E_i=\frac{1-k^4}{8(1+\sqrt{1-k^2})^2}M\,,
\end{equation}
meaning that the semiclassical evaporation formula obtained by Page is valid up to the stage at which $k=a_0/r_0 \lesssim 1- 1/(G_NM^2)^2\approx 1-1/J^2$,  where all mass measurement are in Planck units. Then if we are given the relative ratio of a particle species flux with different spin $s$ to the total flux as $k_s$, ($k_0+k_1+k_2=1$), we can compute the time evolution of the mass parameter as
\begin{equation}
    \frac{d E_i}{dt} =-\sum_{s=0}^{2}k_s \int_0^{E_i} d\omega\Big{(}\frac{d E}{dtd\omega}\Big{)}_{s,l=s,0}\,.
\end{equation}

We can further simplify the above expression using the results obtained in Section \ref{Sec:EmissionRates}. In particular, from Figures \ref{fig:01} and \ref{fig:02} we can roughly estimate that the emission in a channel with angular momentum $l+1$ is twenty orders of magnitude smaller that the emission in the channel for $l=0$. Although 
Figures \ref{fig:01} and \ref{fig:02} pertain to a neutral scalar the result is quite universal for other particles. Regarding particles carrying spin $s$, we find that for $l\geq s$  if a scalar exists, we can almost always neglect photons; if photons exists, we can almost always neglect gravitons and so on. Essentially, the particle with lowest spin $s$ completely dominates the evaporation process. These results are in general agreement with those  obtained by Page in the low frequency limit, for instance, Eq.(19) of \cite{Page:1976ki}.
We now compute the evolution of black hole energy in the microcanonical ensemble with scalar emission. We provide the results of the full evaporation rate for both the semiclassical  and the quantum regimes. In the semiclassical case one has
\begin{equation}\label{microsemi}
   \frac{dE_i}{dt}= -\int_0^{\infty} \Big{(}\frac{d E}{dtd\omega}\Big{)}_{0,l,0}d\omega\approx -T_H^{4\tilde{l}+4}[\pi(r_0^2+a_0^2)]^{2\tilde{l}+1}\Big{|}\frac{\Gamma(\tilde{l}+1)}{\Gamma(\tilde{l}+3/2)}\Big{|}^2f(\tilde{l})\,,
\end{equation}
where 
\begin{equation}
f(\tilde{l})=\int_{0}^{\infty}\prod_{j=1}^{\tilde{l}}(1+(\frac{x}{2\pi j})^2)\frac{x^{2\tilde{l}+3}}{e^{x}-1}dx\,,
\end{equation}
and recalling that in the microcanonical ensemble we have $T_H\sim E_i^{1/2}$, the result is approximately $E_i\sim t^{-1/(2l+1)}$, where $t$ is the time interval from the beginning of evaporation to the point we measure the black hole energy.

\begin{figure}[htbp]
	\centering
\includegraphics[width=0.48\textwidth]{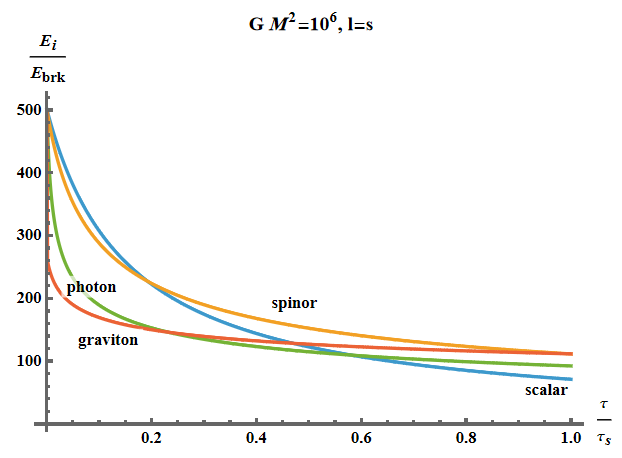}
\includegraphics[width=0.48\textwidth]{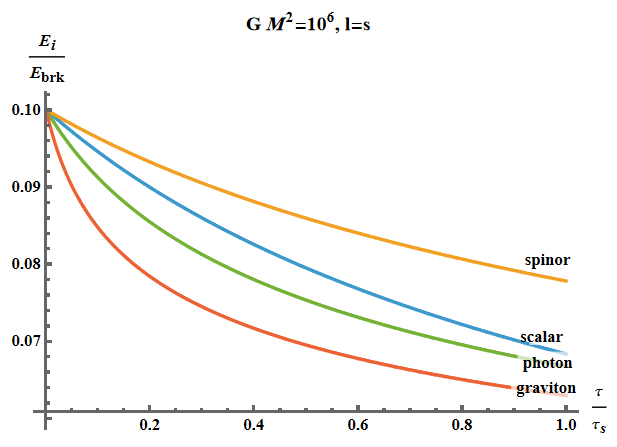}
\caption{The evolution of the energy over extremity with pure scalar/spinor/photon/graviton radiation in microcanonical ensemble. Left panel: with initial value $\zeta_0=(E_i/E_{brk})_{\tau=0}=500$, and we have taken the time unit $\tau_s=10^{21}GE_{brk}\cdot[\frac{5(GM^2)^2}{\zeta_0}]^{2s+1} $ for $s=0,\frac{1}{2},1,2$. Right panel: with initial value $\zeta_0=(E_i/E_{brk})_{\tau=0}=0.1$, and we have taken the time unit $\tau_s=10^{21}GE_{brk}\cdot[\frac{10(GM^2)^2}{\zeta_0}]^{2s+1} $ for $s=0,1,2$. }
\label{fig:06}
\end{figure}

In the quantum regime ($E_i/E_{brk}\rightarrow 0$), approximating $\Delta$ to be an integer leads to,
\begin{equation}
  \Big{(}\frac{d E_i}{dtd\omega}\Big{)}_{0,l,0}\approx -\mathcal{N}_{l0}^2[(\Delta-1)!]^{4}\frac{2\omega^{2\Delta}\sqrt{2C(E_i-\omega)}}{(2C)^{2\Delta-1}\Gamma(2\Delta)}
\end{equation}
or more explicitly 
\begin{equation}\label{full flux}
    \frac{dE_i}{dt}\Big{|}_{\mathrm{micro},E_i}=-\frac{\sqrt{2C}}{4\sqrt{\pi}}\Big{(}\frac{r_0^2+a_0^2}{2C}\Big{)}^{2\tilde{l}+1}(\tilde{l}!)^4\frac{\Gamma(2\tilde{l}+3)}{\Gamma(2\tilde{l}+9/2)}\Big{|}\frac{\Gamma(\tilde{l}+1)}{\Gamma(\tilde{l}+3/2)\Gamma(2\tilde{l}+1)}\Big{|}^2E_i^{2\tilde{l}+7/2}\,.
\end{equation}
The corresponding expression for photons $(s=1)$ and gravitons $(s=2)$, is proportional to the expression of scalar above with the corresponding conformal dimension $\tilde{l}$. We note that in this stage $C=E_{brk}^{-1}$ is almost a constant, and for the leading $s=l,m=0$ channel emission we have $E_i\sim t^{-2/(4l+5)}$. Comparing with the result of semiclassical emission, we find that the decaying rate of the energy over extremity is significantly slower. Away from either regime, we can only compute the total energy flux of evaporation in the microcanonical ensemble numerically. The results are reported in Fig.~\ref{fig:06}.

We can compare this result with other potential channels of evaporation. Comparing the result of single photon emission in Eq.(\ref{full flux}) for $\tilde{l}=1$ (i.e., $l=s=1,m=0$) case, which is proportional to $E_i^{11/2}$ to di-photon (proportional to $E_i^{19/2}$) we find that the latter is much smaller than single-photon emission. In general, it follows from dimensional analysis that di-particle emission with conformal dimension $l$ has an exponent of $4l+\frac{11}{2}$ on $E_i$, which is larger than the single-particle result $2l+\frac{7}{2}$. We thus conclude that the evolution in the quantum regime is governed by single particle emission.

\begin{figure}[htbp]
	\centering
\includegraphics[width=0.48\textwidth]{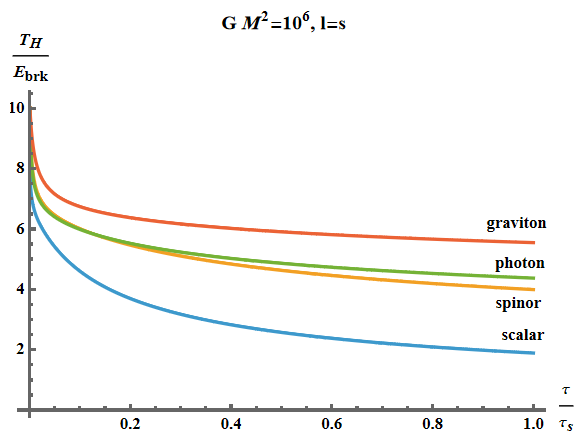}
\includegraphics[width=0.48\textwidth]{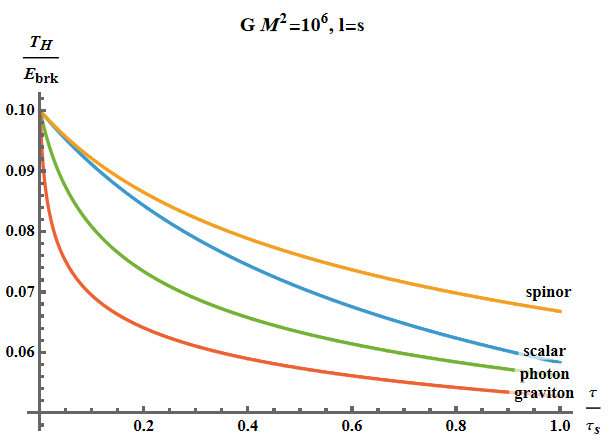}
\caption{The evolution of the temperature with pure scalar/spinor/photon/graviton radiation in canonical ensemble. Left panel:with initial value $\zeta_0=(T_H/E_{brk})_{\tau=0}=10$, and we have taken the time unit $\tau_s=10^{22}GE_{brk}\cdot[\frac{(GM^2)^2}{100\zeta_0}]^{2s+1} $ for $s=0,\frac{1}{2},1,2$.  Right panel: with initial value $\zeta_0=(T_H/E_{brk})_{\tau=0}=0.1$, and we have taken the time unit $\tau_s=10^{21}GE_{brk}\cdot[\frac{(GM^2)^2}{\zeta_0}]^{2s+1} $ for $s=0,1,2$. }
\label{fig:07}
\end{figure}

Since we have obtained the results for the microcanonical ensemble, we can easily convert to the canonical ensemble. Assuming that the black hole remains in a mixed thermal state given by $\rho=Z(\beta)^{-1}\sum_{i}e^{-\beta E_i}\ket{E_i}\bra{E_i}$, we have
\begin{eqnarray}
    \frac{dE_{BH}}{dt}\Big{|}_{\mathrm{canonical}}=\frac{1}{Z(\beta)}\int_0^{\infty}dE_i\rho(E_i)e^{-\beta E_i}\frac{dE}{dt}\Big{|}_{\mathrm{micro},E_i}\,,
\end{eqnarray}
for the average energy of black hole $E_{BH}$. In the large $T_H$ limit the results of different ensembles are identical and we we can directly take Eq.(\ref{microsemi}) as the answer. For the low $T_H$ region, we use Eq.(\ref{full flux}) and find
\begin{eqnarray}\label{evolu}
     \frac{dE_{BH}}{dt}\Big{|}_{\mathrm{canonical}}\approx-\frac{\sqrt{2C}}{4\sqrt{\pi}}\Big{(}\frac{r_0^2+a_0^2}{2C}\Big{)}^{2\tilde{l}+1}(l!)^4(2\tilde{l}+9/2)\Gamma(2\tilde{l}+3)\Big{|}\frac{\Gamma(\tilde{l}+1)}{\Gamma(\tilde{l}+3/2)\Gamma(2\tilde{l}+1)}\Big{|}^2T_H^{2\tilde{l}+7/2}\,,
\end{eqnarray}
indicating that the energy flux in this ensemble is even faster than the semiclassical results. This is not surprising as, via quantum corrections, in the canonical ensemble the relation between energy and temperature has been altered from $E_i\sim T^2$ to $E_i\sim T$. We can also predict the evolution of the temperature and further average energy as LHS of Eq.(\ref{evolu}) is 
\begin{eqnarray}
    \frac{d<E>}{dt}=\frac{d\beta}{dt}[<E>^2-<E^2>]\,.
\end{eqnarray}
We show the results of numerical integration of  the evaporation in the  canonical ensemble  in Fig.\ref{fig:07}. The decay rates in the quantum regime are shown to be much faster.

In conclusion, the quantum corrections hardly change the argument made by Page regarding the dominant channel of emission. There are two main root causes for this result. The first one follows from the fact that the quantum-corrected expression must recover the semiclassical one when $E_i$ is large and, due to the superradiance effect, one has negative $\omega_{\rm eff}$ involved in a way that relaxes the prohibition of certain process with charge exchange involved in the small $E_i$ limit. The second cause arises from the fact that we only consider the degree of freedom in the $z$-direction angular momentum $J_z$ into the ground energy and, therefore,  particles with $l>0$ but $m=0$ may also be allowed. The first reason is a straightforward inference of physical considerations. However, the second one does require some approximations and contain subtleties. In particular, we aim to neglect massive fluctuations in NHEK geometry but on the other hand the $z$-rotation itself cannot satisfy both regularity and being zero modes~\cite{Rakic:2023vhv}. One expects in the low temperature limit that the one-loop effects of quantum gravity are governed by zero modes, leading to a treatment that naturally satisfies the microcanonical ensemble in the quantum regime or the canonical ensemble in low temperature regime. These conditions lead to results that differ from those of~\cite{Brown:2024ajk} which discussed the case of  no charge transmission in the quantum regime and where the emission of photon and gravitons are governed by di-particle emission. Nevertheless, our result that the quantum corrected emission rate in the  microcanonical ensemble is suppressed with respect to the the semiclassical result while in canonical ensemble it is faster is in general agreement with~\cite{Brown:2024ajk}. 

\subsection{Energy and angular momentum emitted}\label{Subsec:evaporation}

We now address the case when the black hole emits not only energy but, more realistically, emits also angular momentum. Our method applies to the case for which as time evolves the black hole gradually approaches a stable and extremal Reissner-Nordstrom configuration with fixed electric charge. The reason why here we basically neglect electric charge emission is motivated by the fact that electrons, the lightest charged particles in the physical spectrum, are all massive and thus their radiation rate is suppressed by exponential decay, as previously noted in~\cite{Brown:2024ajk}. The relevant time scale is $e^{Q\pi G_Nm^2/q}$, in which $q$ and $m$ are the mass and charge of the electron, and therefore, far longer than the time scale of the Hawking radiation we will discuss $t\sim \frac{(G_NM^2)^2}{E_{brk}}$. Hence before that time, the BH will always radiate all its available energy and angular momentum. Let us consider the relation between $q_i$ and $J$. We note that $J\sim kGM^2\gg q_{ex}\approx\sqrt{8\zeta} k$ as $M\rightarrow Q_0$, so $q_i$ can always be regarded as some fluctuation around the $J\rightarrow 0$ regime. 

When we include $U(1)$ emission, generally the dominant contribution to decay of angular momentum comes from the $l=s=m$ single particle emission. The cases with negative $m$ can be neglected because they lead to configurations with $\omega_{eff}=\omega-m\Omega_H\geq \Omega_H$  which are forbidden due to the requirement that conformal dimension in the correlator be real. For these channels, with $m\neq 0$, there should be a distinction between local energy and energy emitted to infinity. For the black hole energy, which is calculated according to Frolov-Thorne vacuum, we have
\begin{eqnarray}
    \frac{dE_i}{dtd\omega}=-2\pi\omega^{2\Delta-1}\omega_{\rm eff}\sum_{l,m}k_{lm}\mathcal{N}_{lm}^2\rho(E_i-\omega_{\rm eff},q_i)|\bra{E_i-\omega_{\rm eff},q_i-m}\mathcal{O}_{\Delta}\ket{E_i,q_i}|^2\,,
\end{eqnarray}
in which $\sum_{l,m}k_{lm}=1$, instead of Eq.(\ref{E infty}). Note that for black holes that spin very fast, in most cases $\omega_{\rm eff}=\omega-m\Omega_H$ is negative. We arrive at the conclusion that the BH energy increases while at the same time there is still positive energy emitted to infinity. The price we pay, however, is the decrease of the angular momentum. The evolution of the angular momentum of the black hole in the microcanonical ensemble is given by 
\begin{equation}\label{ang emitt}
    \frac{dJ}{dt}\Big{|}_{\mathrm{micro},E_i,q_i}=-\sum_{l,m}k_{lm}\int_0^{m \Omega_H+E_i}d\omega\frac{m}{\omega_{eff}}\Big{(}\frac{dE_i}{dtd\omega}\Big{)}_{E_i,q_i,l,m},\,
\end{equation}
while in the  grand canonical ensemble, one has
\begin{eqnarray}
     \frac{dJ}{dt}\Big{|}_{\mathrm{grandcanonical}}=\frac{1}{Z(\beta,\mathcal{E})}\sum_{q_i}\int_{0}^{\infty}dE_i\rho(E_i,q_i)e^{-\beta E_i+2\pi\mathcal{E}q_i}\frac{dJ}{dt}\Big{|}_{\mathrm{micro},E_i,q_i}\,.
\end{eqnarray}

Although in principle we can calculate the emission rate for any channel and any spin with the expression in Eq.(\ref{ang emitt}), technically we meet significant  difficulties for photon and graviton emission in Kerr spacetime. The first significant difficulty pertains to the conformal dimension in the effective formula for the two-point function. The conformal dimension, which is related to the separation variables, cannot be determined analytically and it depends on the frequency region, particularly  when $a_0=r_0$, implying  $r_0\omega\in[0,\frac{m}{2}]$, and the series expansion in Eq.(\ref{ana dim}) breaks down. The second obstacle comes from numerical singularities when applying numerical integration of Gamma functions. The only solution for this second difficulty is to manipulate the original expression into a multiple product form like in Eq.(\ref{factor note}), which requires $\tilde{l}$ to be an integer and this is not generally the case for $a_0=r_0$. Due to these reasons, one can only expect numerical results about the evaporation history of near extremal Kerr black hole, governed by photon and graviton superradiance ($l=m$) channel. The general result is that the evaporation actually will extract the BH away from extremity~\cite{Rakic:2025svg}, in accordance to our argument that in some parameter region the superradiance increase local (relative to extremity) energy. Because the superradiance energy scale is always above the breaking energy, the whole process is controlled by semiclassical formula, in which the energy measured at infinity decay always slower than angular momentum.

As a consequence of these difficulties, our quantitative illustration will focus on scalar emission of slowly rotating Kerr-Newman black hole ($k=a/r_0\ll1$) with available energy $\zeta=E_i/E_{brk}\ll G_NM^2$. In this case, we simply have $\tilde{l}\approx l$. What is interesting about this case is that there are now two competing factors that will influence the energy emission and relative emission rate between energy and angular momentum. First, we notice that for energy emission, $(\delta\omega)^{2l+1}$ leads to s-wave dominance. Second, we should also explore superradiance effects, for example, what is the magnitude of the emission in the  $l\geq m>0$ channel compared to that one of the $l=m=0$ channel? If the former channel is dominant and thus local energy of black hole increases, we will call it the superradiance dominance region, otherwise we have a $s$-wave dominance region in which the black hole energy decreases. In the following we will divide all cases into three types.

\begin{figure}[htbp]
	\centering
\includegraphics[width=0.8\textwidth]{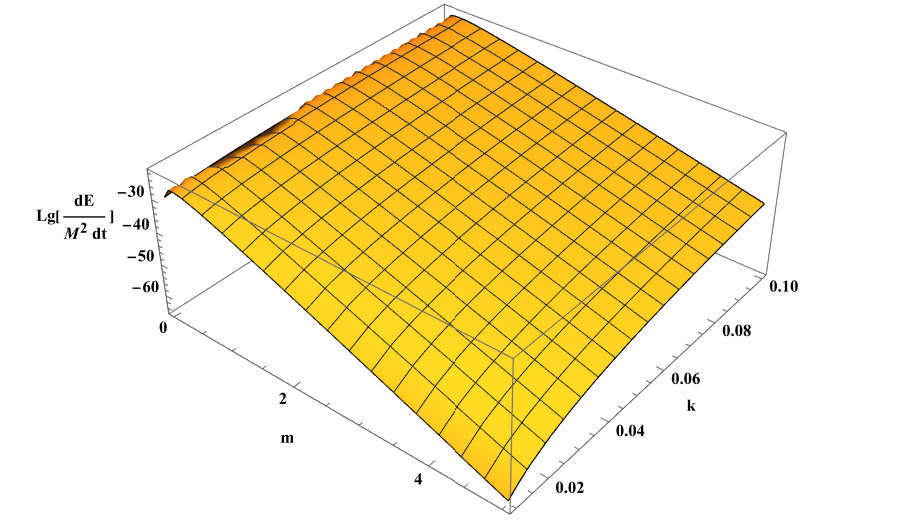}
\caption{The energy flux as a function of quantum number $m$ and spin-mass ratio $k$. }
\label{fig:20}
\end{figure}

The first two types all belong to the assumption that $\Omega_H\gg E_i\gg E_{brk}$ is valid initially which implies that  $k_0\gg \zeta_0/(G_NM^2)$ (from now on all variables with subscript 0 denote the initial value, while the same letter without the subscript stands for the value taken at a certain time, except for $r_0=G_NM$ which is a fixed parameter). We are able to obtain the approximation 
\begin{eqnarray}\label{sup radi}
\begin{aligned}
    \frac{dE_i}{dt}\approx2\pi\mathcal{N}_{lm}^2\int_0^{m\Omega_H}\frac{\omega_{\rm eff}^{2\Delta}(m\Omega_H-\omega_{\rm eff})^{2\Delta-1}}{\Gamma(2\Delta)}d\omega_{\rm eff}\qquad\qquad\qquad\\
    =\frac{1}{2}(m\Omega_H)^2\Big{(}\frac{m^2a^2}{r_0^2+a^2}\Big{)}^{2\Delta-1}\Big{|}\frac{\Gamma(\Delta+2ir_0m\Omega_H)}{\Gamma(\Delta+\frac{1}{2})\Gamma(2\Delta-1)}\Big{|}^2e^{-2\pi r_0m\Omega_H}\frac{\Gamma(2\Delta)\Gamma(2\Delta+1)}{\Gamma(1+4\Delta)}\,,
\end{aligned}
\end{eqnarray}
and if we further focus on the $l=m$ case, we can plot this expression as a 2-variable function of $m$ and $k=a/r_0$ as shown in Fig.~\ref{fig:20}. We note that the peak is basically located at $m\approx 1$ and increases as $k$ increases. Take $k=0.1$ for instance, we find when $l=m=1/2/3$ the energy flux is $dE_i/(M^2dt)\approx 1.26\times 10^{-24}/5.87\times 10^{-29}/1.37\times 10^{-33}$, decreasing four to five or four orders of magnitude respectively. At the same time, for $l=m=0$ we only obtain $dE_i/(M^2dt)\approx -2.1\times 10^{-33}$, which is far smaller than the first two superradiance channel. Under these conditions we conclude that the superradiance effect is dominant and black hole energy will increase. However, when $k=0.01$, the $l=m=1$ channel and the $l=m=0$ channel are already of the same order of magnitude. As we further consider  $k\rightarrow 0$ we will ultimately recover the well-known result that the dominant scalar channel is $l=m=0$, and the black hole energy will decrease. Specifically, by comparing Eq.(\ref{sup radi}) and Eq.(\ref{microsemi}), we find that when $E_i/E_{brk}\gtrsim  (k^2G_NM^2)^2$ the dominant contribution comes from $s$-wave. It seems that once superradiance is dominant it will have a linear growth that is unbounded, however, we  note that there is also angular momentum decay. Suppose the starting point of evaporation to be at $t=0$, when we have already stepped into ``semiclassical" region, {\it i.e.}, $1\ll \zeta_0=E_{i,0}/E_{brk}\ll G_NM^2$ together with spin-mass ratio $k_0$. Then we have two possibilities. 
\begin{enumerate}
    \item If the initial value is $k_0 \lesssim \zeta_0^{1/4}/(G_NM^2)^{1/2}$, (note that this is consistent with the condition $k
_0\gg\zeta_0/(G_NM^2)$ only when $\zeta_0\ll(G_NM^2)^{2/3}$ which is a bit stronger than our starting assumption) we are in the $s$-wave dominance region and have $dE_i/dt\sim -E_i^2$, $E_i\sim t^{-1}$ while $dJ/dt\sim -J^7$, thus for very long times $k\sim J\sim t^{-1/6}$ and we expect to gradually arrive at the superradiance dominance region. However, the detailed picture is slightly different. Explicitly, if we assume  $\theta_0=J_0^4E_{brk}/((G_NM^2)^2E_0)\lesssim 1$ to be the initial value, we find when $E_i$ satisfies
\begin{eqnarray}
    \theta_0^3=\Big{(}\frac{E_i}{E_0}+\gamma\,\theta_0^{3/2}\frac{\zeta_0^{1/2}}{G_NM^2}(1-\frac{E_i}{E_0})\Big{)}^2\frac{E_i}{E_0}\,,
\end{eqnarray}
there is a transition point, at which $k=\zeta^{1/4}/(G_NM^2)^{1/2}$. Here $\gamma$ is generally an ${\cal O}(1)$ number. The second term in the bracket is extremely small within our assumptions, so we neglect it and conclude that when $E_i=E_0\theta_0$ we step into the superradiance dominance region. This requires a time scale 
\begin{eqnarray}\label{ta}
    t_a=\Big{(}\frac{1}{\theta_0}-1\Big{)}\frac{(G_NM^2)^2}{E_0\alpha}\,,
\end{eqnarray}
in which $\alpha$ is also an ${\cal O}(1)$ number. We note that in this time domain $J$ is almost constant, the time scale where $J$ experiences a large decay is
\begin{eqnarray}\label{tb}
    t_b=\frac{(G_NM^2)^3}{\theta_0^{3/2}\zeta_0^{3/2}E_{brk}}\gg t_a\,.
\end{eqnarray}
After $t_a$ we expect a longer time of balance between the $s$-wave and superradiance of the order of magnitude $t_b$, where $dE_i/dt\sim \#J^{8}-\#E_i^2\sim 0$. When $t\gg t_b$, we have exactly $J\sim t^{-1/6}$ and $E_i\sim J^4\sim t^{-2/3}$, thus maintaining an almost balance between the two competing terms, but still slightly and consistently lowering the energy. This is the case until we have $E_i/E_{brk}\lesssim 1$, at which point we need to transfer to the quantum-corrected result for the energy flux. This requires a time
\begin{eqnarray}\label{tc}
    t_c=\frac{(G_NM^2)^3}{E_{brk}}\gg t_b\,,
\end{eqnarray}
and after that we have $dE_i/dt\sim \#J^{8}-\#E_i^{7/2}$, so $E_i\sim t^{-8/21}$. Throughout the whole process $J/E_i$ will be larger and larger thus we have no danger of breaking the assumption $\Omega_H\gg E_i$.

\item If the initial state corresponds to $k_0  \gtrsim \zeta_0^{1/4}/(G_NM^2)^{1/2}$, so $\theta_0 \gtrsim1$, we find $dE_i/dt\sim J^8$ and the energy now has a linear growth within the timescale $t_d$ which is of the same order to $t_a$
\begin{eqnarray}\label{td}
    t_d=\Big{(}1-\frac{1}{\theta_0}\Big{)}\frac{(G_NM^2)^2}{E_0\tilde{\alpha}}\,,
\end{eqnarray}
in which $\tilde{\alpha}$ is an ${\cal O}(1)$ number. We also expect $t_d\ll t_b$ so there will still be a long time balance. When $t\gg t_b$ it has the same asymptotic behavior as to the first situation, $E_i\sim t^{-2/3}$ when $t\gg t_c$ and $E_i\sim t^{-8/21}$ when $t\gg t_c$, with $J\sim t^{-1/6}$ always. 
\end{enumerate}

The last type of initial conditions is when the initial angular momentum is rather small, or equivalently $J_0\ll\zeta_0$, $k_0\ll\zeta_0/(G_NM^2)$ so that $\Omega_H\ll E_i$, which allows to naturally neglect the superradiance effect in any given channel. The dominant channel of energy emission comes from $l=m=0$, while that of angular momentum emission is from $l=m=1$. We then find from Eq.(\ref{microsemi}) that $dE_i/dt\sim- E_i^2$ while $dJ/dt\sim -E_i^{7/2}$. Just like in the first assumption, in the time $t_e$, which is of the same order of $t_a$, $E_i$ is almost a constant and $J$ will have a linear decay, until we have $J=J_0-\zeta_0^{5/2}(G_NM^2)^{-4}$. As long as $J_0/\zeta_0$ is not too small, we can neglect the change of $J$ within this time. When $t\gg t_e$ we have $E_i\sim t^{-1}$ and $J\sim t^{-5/2}$. Finally, we arrive at the quantum regime, which is achieved after the time scale
\begin{eqnarray}\label{tf}
    t_f=\frac{(G_NM^2)^2}{E_{brk}\alpha}\,,
\end{eqnarray}
Then, from Eq.(\ref{full flux}) we have $dE_i/dt\sim -E_i^{7/2}$ and $dJ/dt\sim -E_i^{9/2}$ and thus $E_i\sim t^{-2/5}$, $J\sim t^{-4/5}$. Throughout the process $J/\zeta$ is becoming smaller and smaller and we can always neglect superradiance effects.

\begin{figure}[htbp]
	\centering
\includegraphics[width=0.5\textwidth]{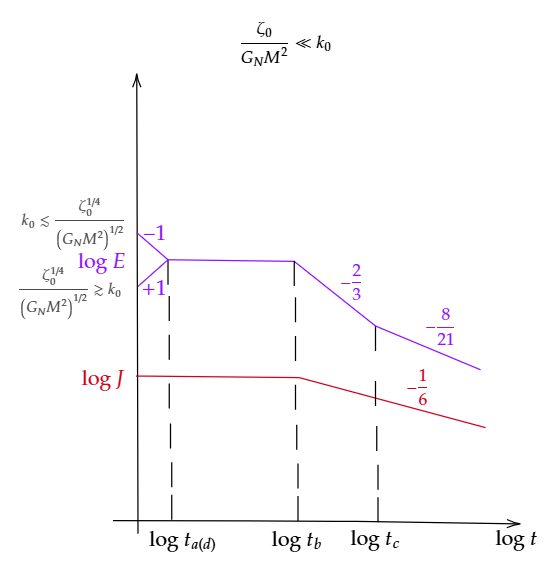}
\includegraphics[width=0.45\textwidth]{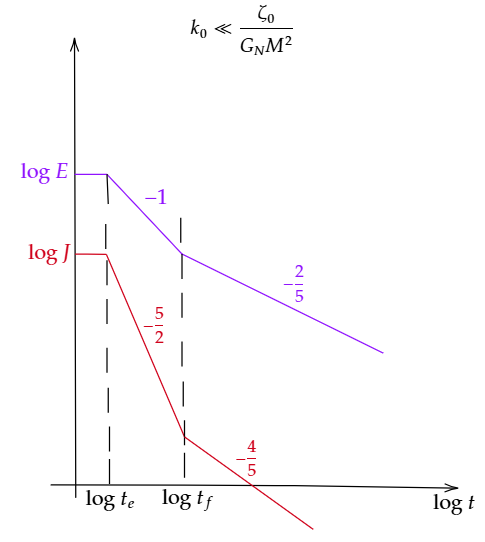}
\caption{Different scenarios of the whole evaporation process. The transition time scale is given by Eq.(\ref{ta}), Eq.(\ref{tb}), Eq.(\ref{tc}), Eq.(\ref{td}) and Eq.(\ref{tf}), the number near the lines are slope and the absolute position of two lines do not  have physical meaning. By comparing them we find in the case when spin is rather large late time evaporation rate is even slower.}
\label{fig:21}
\end{figure}

We sketch the whole evaporation process in Fig.~\ref{fig:21}, including all three possibilities. Our treatment above systematically tracks the fate of the black hole as it begins to approach extremity. There is still a natural question regarding the choice of initial conditions as shown in Fig.~\ref{fig:21}. There are two mutually exclusive situations which we have illustrated. The description on the right panel of Fig.~\ref{fig:21} coincides with the results reported in~\cite{Brown:2024ajk}, which is consistent with the semiclassical calculation that a Kerr-Newman black hole will lose angular momentum faster than its mass and charge. Essentially a near-extremal rotating black hole quickly becomes a near-extremal Reissner-Nordstrom black hole for which the analysis of \cite{Brown:2024ajk} applies. We are left with the interesting puzzle of whether the left panel of Fig.~\ref{fig:21} can actually be realized. 

To address this we note that $k_0\gg \frac{\zeta_0}{G_NM^2}$ combined with the condition we introduce in the classification discussion before, $k_0\sim (G_NM^2)^{-1/3}$, leads to $\zeta_0\sim (G_NM^2)^{2/3}$ which can only occur as an initial state for a black hole with $\zeta_0\sim (G_NM^2)^2 \cdot O(k_0^4)$ whereas $k_0\rightarrow 0$. This implies that the black hole generally spins very slowly, and it behaves very extremal from the very beginning. Even for slowly-spinning black holes, $k_0$ being an ``acceptable" small number generally requires a rather small total mass which is hardly possible for any astrophysically realistic black holes, but still theoretically interesting (for instance, for primordial black holes). The mass cannot be too small, as it will take us outside the regime of validity of the low-energy effective field theory we utilize. Other corrections to  quantum fluctuations, including contributions from other saddles such as wormhole topologies, which are originally suppressed with $e^{-G_NM^2}$, may become comparable to the leading terms we keep. What is more important of this case is that $0<k_0<1$ automatically guarantees that such black holes will be under our high energy cutoff ($\zeta_0\ll G_NM^2$) and thus the effective theory works well from the very beginning. 

For the case described by the right panel, we cannot use the effective theory at once. The quantum corrections discuss in this paper are relevant only for near-extremal black holes. Generically, for the first stages of the evaporation process we assume that the process is semiclassical and adiabatic, the emission rate is predictable, as the temperature, as well as the angular velocity are all determined by the present parameters. Then using the well-known semiclassical results obtained by Page, we are able to control the whole process until it reaches the near-extremality regime:  $M^2-a^2-Q^2<\epsilon$. At this point have passed through the semiclassical boundary and stepped unto the overlapping regime. Then all we need is to alter the semiclassical  expressions by the quantum-corrected ones, and we get conclusions similar to those in \cite{Brown:2024ajk}: this black hole becomes more and more similar to an extremal Reissner-Nordstrom black hole.

\section{Quantum Absorption Cross Section of Near-Extremal Rotating Black Holes}\label{Sec:CrossSection}

The low-energy cross section of black holes is known to admit a fairly universal description. The apex of this universality is that the cross section is proportional to the horizon area in any dimension \cite{Das:1996we}.
Recently, this concept has been revisited in the presence of quantum corrections that arise in the throat region of near-extremal black holes \cite{Emparan:2025sao}. According to its definition, the cross section at a given frequency, $\omega$, can be written as
\begin{eqnarray}\label{formula}
\sigma_{\mathrm{total}}=\frac{2(2l+1)\pi^2}{\omega^2}[\Gamma_{\mathrm{abs}}(\omega)-\Gamma_{\mathrm{em}}(\omega)],
\end{eqnarray}
 where both $\Gamma_{\mathrm{em}}(\omega)$ and $\Gamma_{\mathrm{absorb}}(\omega)$ are transition rates defined as in Eq.(\ref{transition rate}). With these ingredients already  at hand, we address quantum correction to the total cross section for near-extremal rotating black hole.


\subsection{Scalar cross section \& low frequency enhancement}

We start with a warm-up discussion of scalar cross section with only $s$-wave taken into consideration. We will then show certain universality for the results in rotating (possibly charged) black holes. Consider the special case of $l=m=0,\, \Delta=1$ in Eq.~(\ref{formula}) for which we can get an explicit result for the cross section in the microcanonical ensemble
\begin{eqnarray}\label{scalar cross}
\begin{aligned}
    \sigma_{\mathrm{total}}=\frac{4\pi(r_0^2+a_0^2)}{\omega}\cdot [\rho(E_i+\omega)|\bra{E_i+\omega}\mathcal{O}\ket{E_i}|^2-\rho(E_i-\omega)|\bra{E_i-\omega}\mathcal{O}\ket{E_i}|^2]\qquad\\
=A_H\Big{[}\frac{\sinh(2\pi\sqrt{2C(E_i+\omega)})}{\cosh(2\pi\sqrt{2C(E_i+\omega)})-\cosh(2\pi\sqrt{2CE_i})}-\frac{\sinh(2\pi\sqrt{2C(E_i-\omega)})\Theta(E_i-\omega)}{\cosh(2\pi\sqrt{2CE_i})-\cosh(2\pi\sqrt{2C(E_i-\omega)})}\Big{]}\,.
\end{aligned}
\end{eqnarray}

\begin{figure}[htbp]
	\centering
\includegraphics[width=0.7\textwidth]{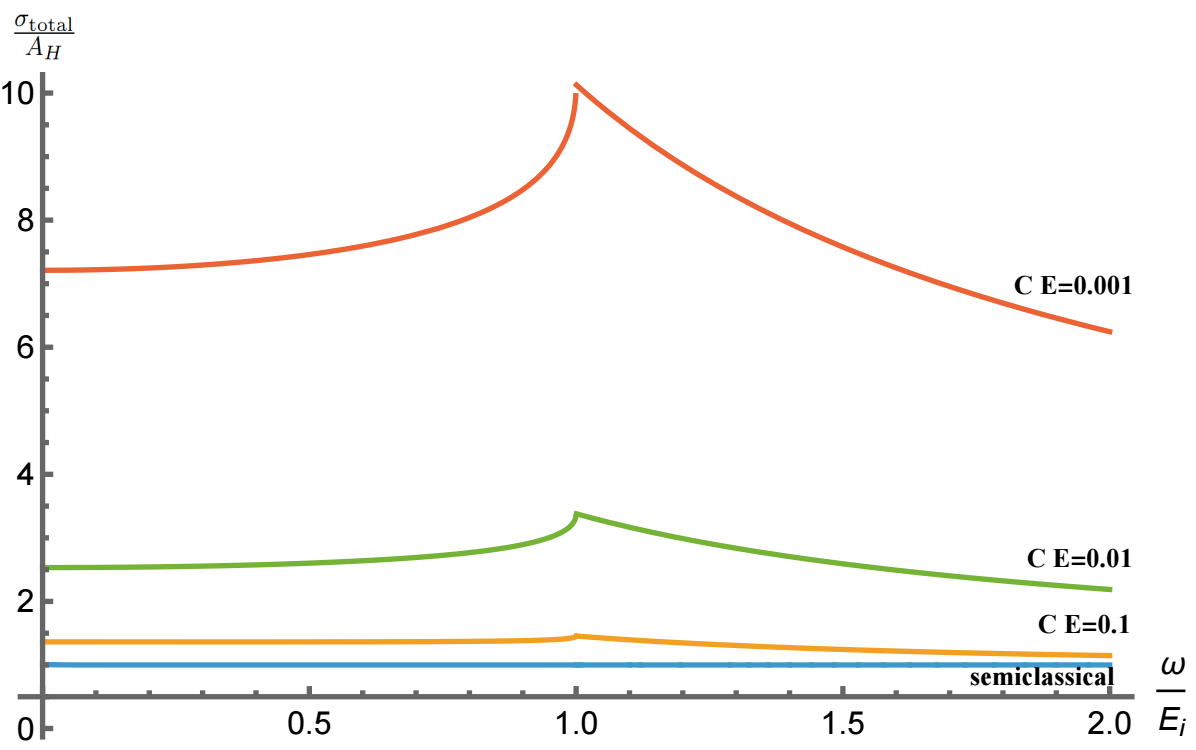}
\caption{The total cross section for $l=0$ scalar, where all three parameter domains are contained. The semiclassical regime corresponds to the blue curve at the bottom and the region near it; the quantum region is at the top; the low frequency limit is on the left side, near the vertical axis.}
\label{fig:10} 
\end{figure}

Let us consider the behavior of the cross section in several different domains in a way that parallels a previous analysis in the spherically symmetric case in \cite{Emparan:2025sao}. In the semiclassical regime, where $E_i/E_{brk}\rightarrow \infty$, we arrive at the expected universal semiclassical result: 
\begin{eqnarray}
    \sigma_{\mathrm{total}}\approx A_H(\frac{1}{1-e^{-\beta\omega}}-\frac{1}{e^{\beta\omega}-1})=A_H\,.
\end{eqnarray}
The universality of this result for spherically symmetric black holes in various dimensions was demonstrated  \cite{Das:1996we}. New behavior arises in the quantum regime, where we set $\omega \ll E_{brk}\,,\; E_i\ll E_{brk}\,,$ the result is
\begin{eqnarray}
     \sigma_{\mathrm{total}}\approx \frac{A_H}{\pi}\Big{(}\sqrt{\frac{E_i+\omega}{2C\omega^2}}-\sqrt{\frac{E_i-\omega}{2C\omega^2}} \Theta(E_i-\omega)\Big{)}\,.
\end{eqnarray}
If we further take the low frequency limit with fixed energy $E_i$, we find
\begin{eqnarray}
\sigma_{\mathrm{total}}\approx A_H\Big{(}\coth(2\pi\sqrt{2CE_i})+\frac{1}{2\pi\sqrt{2CE_i}}\Big{)}\,,
\end{eqnarray}
which is always greater than $A_H$, as first pointed out in \cite{Emparan:2025sao}. The cross section is drastically enhanced in the low energy regime. We present these three different domains together by numerical computations in Fig.~\ref{fig:10}, in which there is a discontinuity at $\omega=E_i$ as expected. As we observe, the universal rule that the $s$-wave sector of cross section is proportional to horizon area up to a quantum correction dimensionless factor still makes sense for quantum rotating black hole just like in the spherical symmetric case. This argument can be made stronger: we prove that the inner structure of the enhancement that relies on frequency is also the same, which can be traced back to the universality of the near horizon geometry and aspects of the low-energy  effective theory. 

The enhancement of the cross section in the low frequency limit shows that although the density of states approaches zero as the energy available above extremality nearly vanishes, the matrix element, which is attained through inverse Fourier transformation of the real-time two-point function $<\mathcal{O}(0)\mathcal{O}(t)>$, must have a greater value than that of the semiclassical one. One can also explain this phenomenon from the correlator point of view, for which \cite{Mertens:2022irh} provides a complete comparison between the quantum and the semiclassical expressions. Indeed, one notes that the behavior of late time 2-pt function goes as $t^{-3}$ which is much greater than the semi-classically expected quasinormal-exponential decay. This regime of the 2-pt function dominates the behavior of low-frequency region of the matrix element $|\bra{E_i+\omega}\mathcal{O}\ket{E_i}|^2$, clarifying the origins of  the enhancement. On the other hand, in the early time regime, $t\rightarrow 0$, there are identical UV divergence behavior, i.e., $<\mathcal{O}(0)\mathcal{O}(t)>\sim t^{-\Delta}$ in the two pictures, and this corresponds to the fact that the quantum cross section presented in Eq.(\ref{scalar cross}) becomes semiclassical in the high frequency regime.

\subsection{Photon and graviton cross section \& quantum transparency}

The discussion and results in the previous subsection proved to be quite similar to the spherically symmetric case, because for that channel no rotating effect can be probed. There are, of course, new features for $l\neq 0$ and $m\neq 0$. The two most prominent ones are the superradiance effect and certain quantum transparency condition. Let us discuss those in more detail. We write Eq.(\ref{cross section}) of these channels as
\begin{equation}\label{cross section}
\begin{aligned}
\sigma_{\mathrm{total}}=\frac{2(2l+1)\pi^2}{\omega^2}\cdot2\pi\sum_{m}\omega^{2\Delta-1}[\mathcal{N}_{lm}^2\rho(E_i+\omega_{\rm eff},q_i+m)|\bra{E_i+\omega_{eff},q_i+m}\mathcal{O}_{\Delta}\ket{E_i,q_i}|^2\\-\mathcal{N}_{l,-m}^2\rho(E_i-\omega_{\rm eff},q_i-m)|\bra{E_i-\omega_{\rm eff},q_i-m}\mathcal{O}_{\Delta}\ket{E_i,q_i}|^2]\equiv \sum_{lm}\sigma_{\mathrm{tot},lm}\,.\qquad\qquad
\end{aligned}
\end{equation}
In the semiclassical regime ($E_i/E_{brk}\rightarrow \infty$) we have 
\begin{eqnarray}
\sigma_{\mathrm{tot},lm}=\frac{(2l+1)\pi}{2^{2\tilde{l}+2}}\omega^{2\tilde{l}-1}\omega_{\rm eff}T_H^{2\tilde{l}}A_H^{2\tilde{l}+1}\Big{|}\frac{\Gamma(\tilde{l}+1+2ir_0\omega)\Gamma(\tilde{l}+1+i\frac{\omega_{\rm eff}}{2\pi T_{H}}-2ir_0\omega)}{\Gamma(\tilde{l}+\frac{3}{2})\Gamma(2\tilde{l}+1)\Gamma(1+i\frac{\omega_{\rm  eff}}{2\pi T_{H}})}\Big{|}^2=\frac{(2l+1)\pi}{\omega^2}\sigma_{lm}\,,
\end{eqnarray}
and one finds it in precise agreement with~\cite{Maldacena_1997}. Note that $\omega_{\rm eff}<0$ corresponds to the superradiance region expressed here as a negative absorption cross section, $\sigma_{{\rm tot}, lm}<0$. There is, of course, no such phenomenon in the spherically symmetric case if one considers neutral particles.

In the quantum regime, $E_i/E_{brk}\rightarrow 0$, the behavior is more involved. Though the idea is similar, the presence of $U(1)$ gauge modes in the density of states makes our discussion of the transparency phenomenon differ from the situation in the spherically symmetric case~\cite{Emparan:2025qqff}. Whether we consider Hawking radiation or its inverse process, absorption of particles, we note that the quantum transition rate is always accompanied by Heaviside theta functions $\Theta(E_f -\frac{q_f^2}{2K})=\Theta(E_i\pm\omega_{eff}-\frac{q_f^2}{4}E_{brk})$, so $\frac{q_f^2}{4}E_{brk}-E_f$ provides a certain lower (upper) bound on the frequency that a black hole can absorb (emit). For some value of the pair $(\omega,m)$, there could emerge a transparency phenomenon, in which the total cross section vanishes; there could also be some discontinuous point at which the radiation or absorption channel opens or closes. We now provide a systematic description of the transparency  phenomenon. 

Let us assume that the quantum numbers $E_i$ and $q_i$ of the initial state are both given and satisfy $E_i-\frac{q_i^2}{4}E_{brk}>0$. Then we have the following conditions (we note that for absorption and emission there is $q_f=q_i\pm m$)
\begin{equation}
    \omega_{eff}>\frac{(q_i+m)^2}{4}E_{brk}-E_i\; \;\;(\mathrm{for\; possible\; absorption})\,,
\end{equation}
\begin{equation}
    \omega_{eff}<E_i-\frac{(q_i-m)^2}{4}E_{brk}\;\;\;(\mathrm{for\; possible\; emission})\,.
\end{equation}
Therefore, for 
\begin{equation}\label{cond1}
    |m|\leq\sqrt{4\zeta-q_i^2}\,,    
\end{equation}
there is no transparency window, as under this condition there is an intersection of these two frequency domains. Note that the two relations above are analogous to the conclusion of transparency frequency in the spherically symmetric black hole case~\cite{Emparan:2025qqff}, where their frequency bound is given by $SU(2)$ energy $\frac{l_f(l_f+1)}{2}E_{brk}-E_f$. However, even when Eq.(\ref{cond1}) is violated, we still cannot assert that there is transparency, as we always require $\omega_{\rm eff}>-m\Omega_H$. Therefore, we also need  $\frac{(q_i+m)^2}{4}E_{brk}-E_i>-m\Omega_H$, which means
\begin{equation}
    m>m_{+} \mathrm{\;or\;} m<m_{-}\,, \; m_{\pm}=-q_i+\frac{K}{4}[-\Omega_H\pm \sqrt{\Omega_H^2+\frac{8}{K}(E_i+q_i\Omega_H})]\,.
\end{equation}
We note that $K\Omega_H=2G_NM^2k\gg \zeta$, implying 
\begin{equation}
    m_{-}\ll - \sqrt{2KE_i-q_i^2}=-\sqrt{4\zeta-q_i^2}\,,
\end{equation}
hence it is nearly impossible to meet this requirement. On the other side, however, we have
\begin{equation}
     m_{+}\approx \frac{E_i}{\Omega_H}=\frac{2\zeta}{K\Omega_H}\ll\sqrt{4\zeta-q_i^2}\,,
\end{equation}
which is satisfied naturally; we find the only possible transparency domain to be $m> \sqrt{4\zeta-q_i^2}$, while $E_i-\frac{(q_i-m)^2}{4}E_{brk}\leq \omega_{\rm eff}\leq \frac{(q_i+m)^2}{4}E_{brk}-E_i$. Otherwise, it is clear that the cross section will vanish only at a point where either $\omega_{\rm eff}=0$ or $\omega=0$ with no discontinuity. Recall that where the effective theory is dominant, the main contribution comes from $m\leq l$ as large as possible and $l$ as small as possible for a given value of the spin $s$. The existence of transparency, therefore, largely relies on whether $\sqrt{4\zeta-q_i^2}$ is an order one number.

\begin{figure}[htbp]
	\centering
\includegraphics[width=0.48\textwidth]{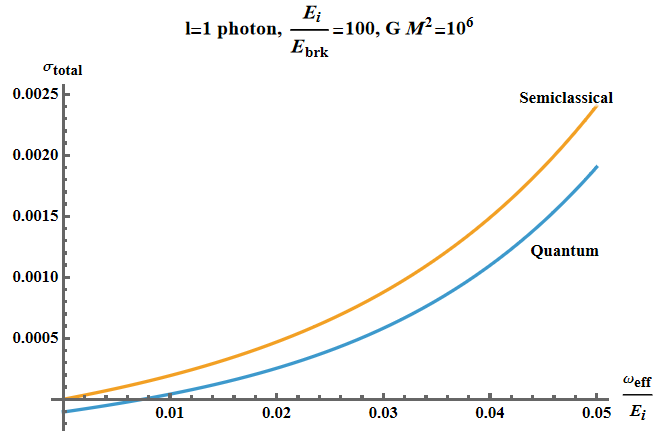}
\includegraphics[width=0.48\textwidth]{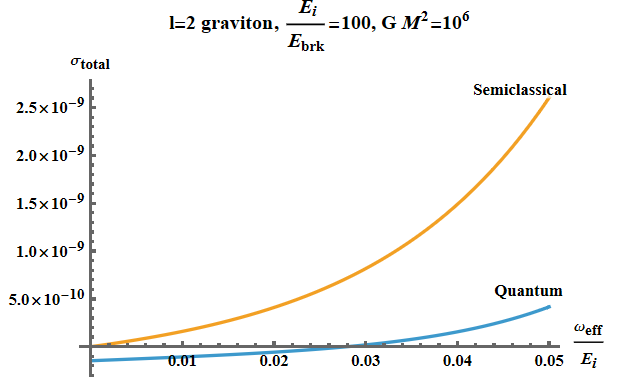}
\caption{The total cross section for $l=1$ photon and $l=2$ graviton, for both semiclassical and quantum results with $\frac{E_i}{E_{brk}}=100$.}
\label{fig:09}
\end{figure}

\begin{figure}[htbp]
	\centering
\includegraphics[width=0.48\textwidth]{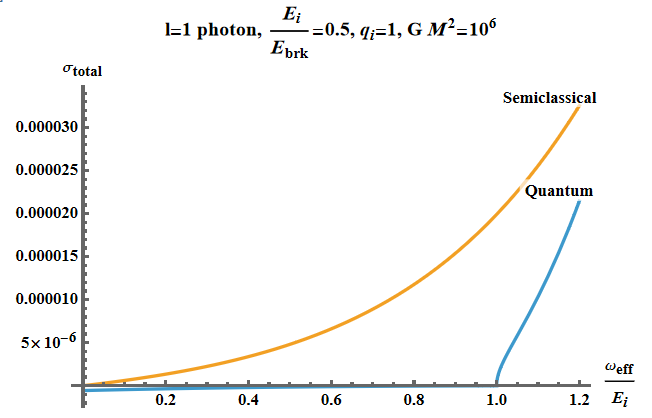}
\includegraphics[width=0.48\textwidth]{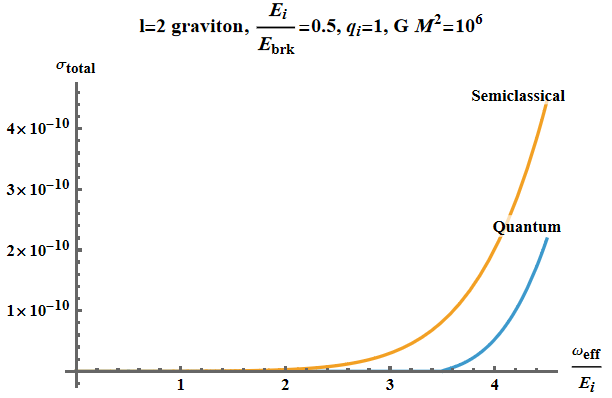}
\caption{The total cross section for $l=1$ photon and $l=2$ graviton, for both semiclassical and quantum results with $\frac{E_i}{E_{brk}}=0.5$ and $q_i=1$.}
\label{fig:19}
\end{figure}

The total cross section for photons and gravitons are shown in Fig.~\ref{fig:09} and Fig.~\ref{fig:19}, respectively. Notably there are some regions where the semiclassical prediction gives positive cross sections while quantum results show negative ones. We conclude that the effective superradiance effect corresponding to negative cross sections appears before the semiclassical prediction. In Fig.~\ref{fig:19} one can clearly see the transparency phenomenon for gravitons as now $m>\sqrt{4\zeta-q_i^2}$, while for photons the parameter has been chosen to satisfy $m=\sqrt{4\zeta-q_i^2}$, in which case there is no transparency but at the critical point $\sigma=0$ there is a discontinuity, as expected.

Although in Fig.~\ref {fig:09} and Fig.~\ref{fig:19} the quantum cross sections seem to always lie below the semiclassical ones, for high frequencies, generally the former will first exceed the latter and finally they become identical. Here we mainly present the low frequency domain because it contains more interesting results. This behavior is quite different from  what was verified in the $s$-wave channel, in which at low frequencies the quantum cross section is always higher than the semiclassical one while at high frequencies they become identical. As we have mentioned, although we sum over all possible values of $m$ for a given  fixed value of $l$ in the definition of the cross section, we still observe approximate transparency as the influence of lower channels is negligible. 

\begin{figure}[htbp]
	\centering
\includegraphics[width=0.48\textwidth]{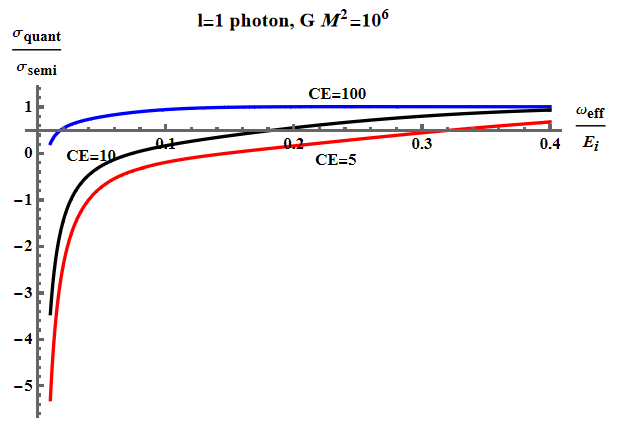}
\includegraphics[width=0.48\textwidth]{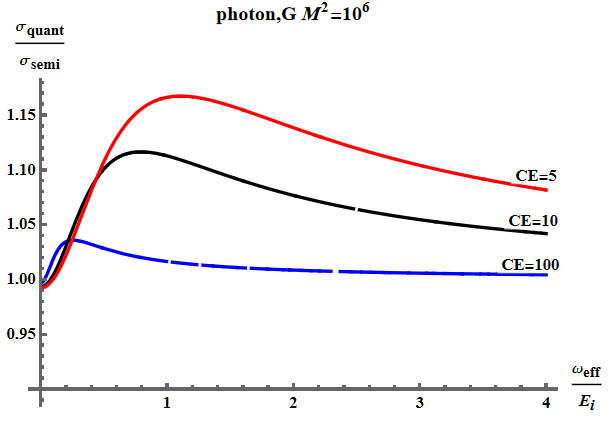}
\caption{The ratio of cross section prediction by quantum computation and semiclassical for photon in rotating spacetime (left) and non-rotating spacetime (right), with different initial energy.}
\label{fig:29}
\end{figure}

We show the comparison of quantum-semiclassical cross section ratio, $\frac{\sigma_{\mathrm{quant}}}{\sigma_{\mathrm{semi}}}$, between spherically symmetric and rotating background in Fig.~\ref{fig:29}, in which for the rotating ones  we have highlighted the behavior in the low frequency region. The main distinction from that of the spherically symmetric case is the existence of negative quantum cross sections, and we expect the ratio to approach $-\infty$ as $\omega_{\rm eff}\rightarrow0$; this phenomenon, resembling superradiance, is exclusive to rotating black holes. For the spherically symmetric case the ratio turns out to be identity for various different initial energy when taking low frequency limit, see right panel of Fig.~\ref{fig:29}. However, we note that high frequency universality introduced in previous subsection makes sense for both rotating and non-rotating cases.

\section{Conclusion and Discussion}\label{Sec:Discussion}
In this manuscript we studied quantum corrections to the emission rate and evaporation process in near-extremal rotating black holes. We implemented the approach of applying the Schwarzian quantum mechanics framework to the study of strongly coupled quantum gravity effect in the near-horizon throat region. The most notable deviation from that of spherically symmetric case~\cite{Brown:2024ajk, Lin:2025woff} is the introduction of a gauge field in the dimensional reduction to describe gauge modes that govern the fluctuation of quantum gravity when reducing from $SO(3)$ symmetric to $U(1)$ axi-symmetric situations. In the case of rotating black holes, the direct approach of dimensionally reducing the problem to two dimensions is not available as opposed to the case in the spherically symmetric case~\cite{Iliesiu:2020qvm,Mertens:2019tcm}. Therefore, rather than dimensionally reducing the problem, we take a more low-energy effective action approach that includes the Schwarzian action and the action of certain gauge modes taking a parallel approach to \cite{Davison:2016ngz}. We studied the compatibility of the quantum framework with the semi-classical approach and found an important link between angular momentum and electric charge exchange.  The natural requirement that the quantum result can recover the semiclassical one in a certain limit is compatible with charge conservation during the evaporation. This is achieved through  alteration of the approximation condition as well as the semiclassical greybody factor, accompanied by taking the proper ensemble on the quantum side to guarantee a match.  

If we summarize the main new ingredient in the quantum treatment of Hawking radiation implemented in \cite{Brown:2024ajk} as imposing the constraint due to  the quantum correlators, namely, a quantum black hole cannot emit more energy than it has, which leads to a vanishing of the emission in certain regime that is allowed semiclassically. Then,  in our case we have the same constraint enhanced by the fact that there are two U(1) charges: angular momentum and electric charge. Now the correlator imposes relations between exchange of angular momentum and electric charge, this is the new effect that we systematically explore in this manuscript. We develop this intuition technically through a number of approaches. From the point of view of the saddle point approximation  in Subsection \ref{Subsec:EffectiveQTheory} (see equations \eqref{saddle}-\eqref{Eq:width}); from  discussing the correlator expression in Subsection \ref{Subsec:Coupling} (see equation \eqref{Eq:exchange}); by connecting the quantum expression to the expected semiclassical emission in Subsection \ref{scalar} (equations \eqref{cen2}-\eqref{Eq:charges-initial-final}, verified through evaluation in equation \eqref{factor note}). The implications for emission, total evaporation and absorption are worked out numerically in Sections \ref{Sec:Evaporation} and \ref{Sec:CrossSection}.

With the precisely formulated  quantum gravitational approach at hand,  we were able to compute the emission rate and evaporation process of near-extremal rotating black holes. We presented the result for both the $m=0$ and the $m\neq 0$ channels and highlighted some differences in their behavior. When discussing the whole process of evaporation of charged and rotating black holes, we assumed that the Schwinger pair effect can be neglected and thus the electric charge was kept fixed. We first analyzed the case where only energy is emitted and then the case where the black hole emits  angular momentum as well as the energy above extremity  simultaneously. The latter case is not only more physically relevant but also more interesting. We observe a competing mechanism between the $s$-wave channel and the superradiance channel, and we conclude that in the final stage the two must arrive at a sort of  balance, leading to the energy being emitted at slower than originally expected rate. We also studied the quantum cross section and observe new features including an  enlargement of the superradiance region, quantum transparency and a low energy enhancement effect.  

For future studies, it would be interesting to generalize the Kerr-Newman case we discuss here to  richer physical situations such as Kerr-Newman-dS evaporation which has added difficulties brought about by  issues of cosmological horizons and their effects on the dynamics  \cite{Bhattacharjee:2025wfv}. One could also consider other kinds of type-D extremal spacetimes in possibly higher dimension which have already been successfully discussed via dimensional reduction to the JT model \cite{Castro:2018ffi,Moitra:2019bub,Poojary_2023,Castro_2021,Luo_2025}, including supersymmetric black holes \cite{Lin:2025woff}. It would also be promising to study the attractor mechanism of energy distribution evolution, as a natural extension of recent interesting work reported in~\cite{Emparan:2025sao,Biggs:2025nzs,Emparan:2025qqf,Betzios:2025sct} to the rotating case.

The prospect of directly connecting properties of quantum black holes to aspects of quantum fluids at very low temperatures via the fluid/gravity correspondence  is particularly tantalizing given recent explorations in \cite{Nian:2025oei,PandoZayas:2025snm,Cremonini:2025yqe,Gouteraux:2025exs,Kanargias:2025vul}.  It would be quite interesting to exploit this setup to gain a better conceptual footing on hydrodynamics in the very low temperature regime where typical notions, such as mean free path, are challenged.  Perhaps more audaciously would be to investigate the implications of the effects discussed in this manuscript in the context of more phenomenological questions. Some particularly interesting examples are  the superradiance instability of Kerr black holes as a source for axion searches  \cite{Arvanitaki:2014wva} or various scenarios for primordial black holes \cite{Guo:2017njn}.

\section*{Acknowledgments}
We are thankful to Giullio Bonelli, Roberto Empar\'an, Blaise Gout\'eraux, Xiao-Long Liu, Sabyasachi Maulik, Xin Meng, Jun Nian, Alessandro Tanzini, Jun-Kai Wang, Cong-Yuan Yue, Jingchao Zhang and Yue Zhao for clarifying discussions. This work is partially supported by the U.S. Department of Energy under grant DE-SC0007859.

\bibliographystyle{JHEP}
\bibliography{Hawk-Kerr}

\end{document}